\documentclass[11pt,a4paper]{article}
\usepackage{jcappub}
\usepackage{bm}
\usepackage{physics}
\usepackage{bbm}
\usepackage{comment}
\usepackage{amsmath}
\usepackage{etoolbox}
\usepackage{float}
\usepackage{empheq}
\usepackage{eqparbox}
\usepackage{slashed}
\usepackage{marginnote}

\renewcommand\[{\left[}

 \usepackage{comment} 
 \usepackage[dvipsnames]{xcolor}
\usepackage{float}
\usepackage{amsmath}
\usepackage{empheq}
\usepackage[theorems,skins]{tcolorbox}
\usepackage{letltxmacro}
\newtcolorbox{mymathbox}[1][]{colback=white, sharp corners, #1}
\newtcolorbox{mybox}[2][]{
sharp corners,
colframe=blue!20!black,
  colframe=blue!20!black,
  fonttitle=\bfseries,
  colbacktitle=blue!20!black,
  enhanced,
  attach boxed title to top center={yshift=-2mm},
  title={#2},#1}

\newcommand{\bP}{{\bf P}}

\newcommand{\bp}{{\bf p}}
\newcommand{\be}{{\bm\epsilon}}

\newcommand{\bR}{{\bf R}}
\newcommand{\br}{{\bf r}}
\newcommand{\bv}{{\bf v}}
\newcommand{\bk}{{\bf k}}
\newcommand{\bq}{{\bf q}}

\newcommand{\ri}{{\rm i}}
\newcommand{\rf}{{\rm f}}
\newcommand{\rs}{{\rm s}}

\long\def\exclude#1{}

\usepackage{etoolbox}

\makeatletter
\gdef\@fpheader{}
\makeatother

\begin{document}

\hfill CERN-TH-2023-035

\makeatletter
\makeatother

\title{Stellar limits on
scalars from electron-nucleus
bremsstrahlung}

\author[a]{Salvatore Bottaro,}
\author[b]{Andrea Caputo,}
\author[c]{Georg Raffelt,}
\author[d]{\hbox{and Edoardo Vitagliano}}

\affiliation[a]{School of Physics and Astronomy, Tel-Aviv University, \\ Tel-Aviv 69978, Israel}
\affiliation[b]{Theoretical Physics Department, CERN, 1211 Geneva 23, Switzerland}
\affiliation[c]{Max-Planck-Institut f\"ur Physik (Werner-Heisenberg-Institut), \\F\"ohringer Ring 6, 80805 München, Germany}
\affiliation[d]{Racah Institute of Physics, Hebrew University of Jerusalem, Jerusalem 91904, Israel}

\emailAdd{salvatoreb@tauex.tau.ac.il}
\emailAdd{andrea.caputo@cern.ch}
\emailAdd{raffelt@mpp.mpg.de}
\emailAdd{edoardo.vitagliano@mail.huji.ac.il}

\abstract{We revisit stellar energy-loss bounds on the Yukawa couplings $g_{\rm B,L}$ of baryophilic and leptophilic scalars $\phi$. The white-dwarf luminosity function yields $g_{\rm B}\lesssim 7 \times 10^{-13}$ and \hbox{$g_{\rm L}\lesssim 4 \times 10^{-16}$}, based on bremsstrahlung from ${}^{12}{\rm C}$ and ${}^{16}{\rm O}$ collisions with electrons. In models with a Higgs portal, this also implies a bound on the scalar-Higgs mixing angle \hbox{$\sin \theta \lesssim 2 \times 10^{-10}$}. Our new bounds apply for $m_\phi\lesssim {\rm 1~keV}$ and are among the most restrictive ones, whereas for $m_\phi\lesssim 0.5\,{\rm eV}$, long-range force measurements dominate. Besides a detailed calculation of the bremsstrahlung rate for degenerate and semi-relativistic electrons, we prove with a simple argument that non-relativistic bremsstrahlung by the heavy partner is suppressed relative to that by the light one by their squared-mass ratio. This large reduction was overlooked in previous much stronger bounds on $g_{\rm B}$. In an Appendix, we provide fitting formulas (few percent precision) for the bremsstrahlung emission of baryophilic and leptophilic scalars as well as axions for white-dwarf conditions, i.e., degenerate, semi-relativistic electrons and ion-ion correlations in the ``liquid'' phase.
}

\maketitle

\section{Introduction}

The emission of radiation through bremsstrahlung, due to the deceleration of a particle when deflected by another one, is a staple of classical and quantum field theory. The radiated power was derived by Larmor in 1897 \cite{Larmor:1897}, the quantum-mechanical problem was first solved by Sommerfeld in 1931~\cite{https://doi.org/10.1002/andp.19314030302}, and numerical solutions were analyzed in a seminal paper by Karzas and Latter in 1961 \cite{1961ApJS....6..167K}. While the problem of bremsstrahlung emission could appear quaint, the topic has been revisited several times over the years. The full relativistic cross section was found only in 1969~\cite{1969PhRv..183...90E}. In recent years, the numerical results have been updated~\cite{2014MNRAS.444..420V,2015MNRAS.449.2112V}, and new approximate formulae were obtained~\cite{Pradler_2021}.
Particles other than photons can be emitted as well. In an astrophysical plasma, electron-proton bremsstrahlung can copiously produce neutrino pairs, as first proposed in Refs.~\cite{Pontecorvo:1959wb,Gandelman:1960ex}. Neutrino bremsstrahlung is the dominant neutrino emission process in stars with low temperature and high electron density~\cite{Cazzola:1971ru,Dicus:1976rj}. The emission from a non-degenerate, non-relativistic plasma was obtained again in Ref.~\cite{Vitagliano:2017odj}. Electron-proton bremsstrahlung in the Sun provides the largest keV-range neutrino flux at Earth \cite{Vitagliano:2019yzm}.

The neutrino-proton bremsstrahlung process $\nu+p\rightarrow \nu+p+\gamma$ has been proposed as an  ambitious approach to measure neutrino masses and to distinguish their Dirac vs.\ Majorana nature by looking at the kinematic endpoint~\cite{Millar:2018hkv}. Moreover, putative particles beyond the standard model could interact with electrons and nucleons. The emission of photons in dark matter-proton bremsstrahlung processes has been proposed as a detection channel for sub-GeV dark matter searches~\cite{Kouvaris:2016afs,Bell:2019egg}.
These examples are cases of ``inverted kinematics'' in the sense that the collision is between a heavy and a light particle, where the energy comes from the light partner, whereas the radiation is emitted by the heavy one. We will pay careful attention to answer precisely this question: how is bremsstrahlung modified by inverting the roles of who provides the energy and who radiates in a collision.

In stellar plasmas, feebly interacting bosons instead of neutrino pairs can be produced through bremsstrahlung as well \cite{Raffelt:1996wa}. For example, axions coupling to electrons~\cite{Krauss:1984gm,Raffelt:1985nk,Redondo:2013wwa} are produced in $ee$ or $eN$
collisions. The emission rates of axions and photons are related to each other, with the former being suppressed compared to the latter by a factor $\mathcal{O}(\omega^2/m_e^2)$
in the non-relativistic limit where $\omega\ll m_e$ \cite{Redondo:2013wwa}. Perhaps surprisingly, the vector and axial currents contribute at the same level to neutrino bremsstrahlung~\cite{Vitagliano:2017odj}, so one should be careful in the parametric rescaling of the results concerning different couplings. Moreover, non-relativistic expansions should be handled with care (compare e.g.\ the results of Refs.~\cite{Vitagliano:2017odj,Gandelman:1960ex} in the context of neutrinos, and \cite{Dimopoulos:1986mi,Pospelov:2008jk} in the context of axions).

Novel CP-even bosons that couple to ordinary matter can be copiously produced in stars. Scalar production through bremsstrahlung in a non-relativistic plasma was considered in Ref.~\cite{Grifols:1988fv} and revisited in Ref.~\cite{Hardy:2016kme}, where it was shown that they can be produced through resonant conversion of longitudinal plasmons. Electrophilic scalars with a coupling $g_\phi \phi \Bar{e}e$ are mostly produced in this way. Nucleophilic scalars $g_\phi \phi \Bar{N}N$ can be emitted by this process as well with a proton thermal loop (spectator). Parametric estimates suggested that the strongest constraint on $g_\phi$ came from the evolution of red giants~\cite{Hardy:2016kme}. Particles coupling to both nucleons and electrons in a Higgs-portal fashion, $(m_f/v)\sin{\theta} \, \phi \bar{f}f$ with $v=246\,\rm GeV$, can emerge from this process with both protons and electrons as bystanders. The largest contribution to the scalar production would come from the electron coupling, providing a bound $\sin{\theta}\lesssim 3\times 10^{-10}$ for $m_\phi\lesssim 1 \,\rm keV$ based on the evolution of red giants \cite{Hardy:2016kme}.

Subsequently, other authors found that Higgs-portal scalars were mostly produced in electron-nucleus bremsstrahlung, with the nucleus radiating the particle~\cite{Dev:2020jkh,Balaji:2022noj}. The kinematics is somewhat peculiar: a light particle shakes a heavy one, which in turn emits the radiation. White-dwarf cooling then implied a bound $\sin{\theta}\lesssim 10^{-17}$ \cite{Dev:2020jkh, Balaji:2022noj}, many orders of magnitude more stringent than the previous one. If true, this powerful constraint would strongly impact the parameter space of scalars mixing with the Higgs~\cite{Fradette:2018hhl}, such as in relaxion models~\cite{Flacke:2016szy, Chatrchyan:2022pcb, Chatrchyan:2022dpy}, and suggests that white dwarfs can probe the CP-violating scalar coupling of QCD axions originating in the weak sector of the standard model~\cite{OHare:2020wah}.

Unfortunately, this amazing constraint is suspicious because the alleged bremsstrahlung rate is not suppressed by the factor $(m_e/m_N)^2$ that one would expect for dipole radiation and that was explicitly found in the classical Larmor formula for scalars~\cite{Ren:1993bs}\footnote{We acknowledge a correspondence with the authors of Refs.~\cite{Dev:2020jkh,Balaji:2022noj}, in which they unfortunately dismissed our concern and reaffirmed their conclusion that the squared nucleon mass would {\em not\/} appear in the denominator of the bremsstrahlung emission rate for baryophilic scalars. This and other discrepancies are discussed at the end of Sec. \ref{subsec:bounds}.}. 

Motivated by the important consequences of this question, we revisit the emission of baryophilic and leptophilic scalars in electron-nucleus bremsstrahlung. We obtain explicit expressions for the energy-loss rate and confirm the expected $(m_e/m_N)^2$ suppression relative to the results of Refs.~\cite{Dev:2020jkh,Balaji:2022noj}. We obtain analytical results for a non-relativistic, non-degenerate plasma and for the degenerate case, showing that bremsstrahlung is of course strongly suppressed by Pauli blocking, contrary to previous findings. Moreover, for the conditions of a white-dwarf interior, we properly treat the strong ion-ion correlations and provide simple and accurate fitting formulas.

The rest of the paper is organized as follows. In section~\ref{Scalar_brem} we show with simple arguments the expected scaling of baryophilic vs.\ leptophilic scalar bremsstrahlung radiation, i.e.\ how the mass of the radiating particle in a collision affects the rate. In section~\ref{Sec:el_proton} we find the emission rates, including screening effects for different conditions. In section~\ref{sec:bounds} we derive bounds on novel scalars from the white-dwarf luminosity function in analogy to earlier studies for axions. Finally, section~\ref{sec:conclusions} is dedicated to a summary and discussion. Several technical issues are relegated to appendices. In particular, we develop fitting formulas for the emission of baryophilic and leptophilic scalars as well as axions for white-dwarf conditions. The formulas for axions are somewhat more precise than earlier ones in the literature.

\section{Scalar bremsstrahlung: Quantum mechanics and classical limit}
\label{Scalar_brem}

In this section, we consider non-relativistic bremsstrahlung in electron-nucleus collisions using time-dependent perturbation theory in quantum mechanics and also using the classical limit. We show that the emission of scalars is perfectly analogous to that of vectors (photons) except for the different number of polarization states. In both cases, in a collision or in an atomic transition, the radiation emitted by the heavy partner (e.g.\ the proton in a hydrogen atom), is suppressed by an approximate factor $(m_e/m_p)^2$ that was overlooked in previous studies of baryophilic scalar bremsstrahlung \cite{Dev:2020jkh, Balaji:2022noj}. Our elementary reasoning supports the same finding in a detailed quantum-field theory calculation in Section~\ref{Sec:el_proton}, although in the semi-relativistic plasma of a white dwarf, there are small corrections to this simple scaling.

\subsection{Photon radiation in quantum mechanics}

Let us start considering the emission of photons by two interacting charged particles. This includes bremsstrahlung (free-free emission), but also free-bound or bound-bound processes, the latter equivalent to atomic transitions. Following textbook discussions (e.g.\ Weinberg \cite{weinberg_2015}) and also a recent detailed study of quadrupole radiation \cite{Pradler:2020znn}, we consider two particles with masses $m_{n}$, $n=1$ or~2, and electric charges $e_{n}=Z_n e$ with $e$ the positive unit of electric charge, defining the fine-structure constant as $\alpha=e^2/4\pi$. The initial and final two-particle states $|\ri\rangle$ and $|\rf\rangle$ are assumed to be eigenstates with energies $E_{\ri,\rf}$
of the Hamiltonian
\begin{equation}\label{eq:H0}
  H_0=\frac{\bp_1^2}{2m_1}+\frac{\bp_2^2}{2m_2}+V(\br),
\end{equation}
where $\br=\br_1-\br_2$ is the relative coordinate between the
particles and $V(\br)$ a central potential, here essentially a screened Coulomb potential.

The non-relativistic interaction Hamiltonian of charged particles with photons in Cou\-lomb gauge is $-(e_n/m_n)\,\bp_n\cdot{\bf A}(\br_n)$, where ${\bf A}(\br_n)$ is the photon vector potential in the interaction picture at location $\br_n$ of the particle $n=1$ or 2. The matrix element of the interacting-particle states for the emission of a photon is therefore
\begin{equation}\label{eq:Matrixelement-Photon}
  {\cal M}_{\rf\ri}=-\be\cdot\Bigl\langle\rf\,\Big|
  \frac{e_1}{m_1}\,\bp_1e^{-i\bq\cdot\br_1}+\frac{e_2}{m_2}\,\bp_2e^{-i\bq\cdot\br_2}
  \Big|\,\ri\Bigr\rangle,
\end{equation}
where $\be$ is the real photon polarization vector (i.e.\ describing linear polarization states) and $\bq$ its momentum.

Only the relative motion of the two particles, not the CM motion, can
lead to radiation and so one uses the CM coordinates $\bP=\bp_1+\bp_2$, $\bR=(m_1\br_1+m_2\br_2)/M$, $\br=\br_1-\br_2$, and $\bp=(m_2\bp_1-m_1\bp_2)/M$ with $M=m_1+m_2$ the total mass. One easily confirms that $\bR$ and $\bP$ as well as $\br$ and $\bp$ fulfill canonical commutation relations, whereas $\br$ and $\bP$ as well as $\bR$ and $\bp$ commute. The reverse mapping is $\bp_1=m_1\bP/M+\bp$,
$\bp_2=m_2\bP/M-\bp$, $\br_1=\bR+m_2\br/M$, and $\br_2=\bR-m_1\br/M$. In the new canonical variables, the particle Hamiltonian is
\begin{equation}\label{eq:H0-2}
  H_0=\frac{\bP^2}{2M}+\frac{\bp^2}{2m}+V(\br),
  \quad\hbox{where}\quad
  m=\frac{m_1m_2}{m_1+m_2}
\end{equation}
is the reduced mass. The operator sandwiched between $|\ri\rangle$ and $|\rf\rangle$ in Eq.~\eqref{eq:Matrixelement-Photon} reads in the new variables
\begin{equation}\label{eq:Matrixelement-Photon-2}
  \biggl[\frac{\bP}{M}\Bigl(e_1\,e^{-i\bk\cdot\br\,m_2/M}
  +e_2\,e^{i\bk\cdot\br\,m_1/M}\Bigr)
  +\bp\Bigl(\frac{e_1}{m_1}\,e^{-i\bk\cdot\br\,m_2/M}
  -\frac{e_2}{m_2}\,e^{i\bk\cdot\br\,m_1/M} \Bigr)\biggr]e^{-i\bk\cdot\bR}.
\end{equation}
Following the textbook literature (e.g.\ Weinberg Sec.~11.7 \cite{weinberg_2015}) we notice that the factor $e^{-i\bq\cdot\bR}$ introduces a recoil on the radiating system by the momentum $\bq$ of the emitted radiation, which we neglect in the ``long wavelength approximation'' where $\bq$ is much smaller than the momenta of the radiating particles. In a bremsstrahlung process, this means to neglect $\bq$ in the momentum-conserving $\delta$ function. In a free-bound or bound-bound transition, it means to ignore the recoil of the final-state bound object. In the CM frame, we may also ignore the term proportional to $\bP$.

In the remaining term, we expand the exponentials up to first order, finally leading to the matrix element in the CM frame
\begin{equation}\label{eq:Matrixelement-Photon-3}
  {\cal M}_{\rf\ri}=-\underbrace{m\biggl(\frac{e_1}{m_1}-\frac{e_2}{m_2}\biggr)}_{\textstyle{\cal A}_1}
  \frac{\langle\rf|\be\cdot\bp|\ri\rangle}{m}
  +i\underbrace{m^2\biggl(\frac{e_1}{m_1^2}+\frac{e_2}{m_2^2}\biggr)}_{\textstyle{\cal A}_2}
  \frac{\langle\rf|(\be\cdot\bp)\,(\bq\cdot\br)|\ri\rangle}{m}.
\end{equation}
The first term represents dipole (E1) radiation, whereas the second one corresponds to quadrupole (E2) radiation as well as magnetic dipole (M1) radiation, the latter related to orbital angular momentum. We have not included possible magnetic dipoles of the charged particles that would also contribute on that order if the magnetic moment is roughly that of a Dirac fermion. Actually, a spin-flip transition can be the dominant effect as e.g.\ in the 21~cm hyperfine transition in hydrogen or the 14.4~keV nuclear transition in ${}^{57}{\rm Fe}$ that has been used in solar axion searches \cite{Haxton:1991pu, CAST:2009jdc, Derbin:2011zz, CUORE:2012ymr}. However, we are primarily interested in the emission of scalars where such effects do not occur, in contrast to pseudoscalars such as axions.

For normal atomic transitions or electrons colliding with charged particles $(Z,A)$ in a stellar plasma, we have $m_1=m_e$, $e_1=e$, $m_2=A m_N$ (nucleon mass $m_N$), and $e_2=Z e$, implying $e_1/m_1\gg e_2/m_2$ and the reduced mass $m\simeq m_e$ so that $|{\cal A}_1|\simeq|{\cal A}_2|\simeq e$. For electron-electron collisions, the dipole term vanishes and the quadrupole term dominates~\cite{Pradler:2020znn}. In this case one may say that the center-of-mass and the center-of-charge coincide so that there is no time-changing electric dipole moment that is needed to emit radiation. However, the dipole term dominates unless it cancels for particles with equal $e/m$ or unless in an atomic transition it is forbidden by the quantum numbers of the participating atomic states.

More specifically, the quadrupole term is the next order in an expansion in $\mathbf{q\cdot}\mathbf{r_n}$. That is to say, the quadrupole operator is suppressed with respect to the dipole by a factor $\mathbf{q}/\mathbf{p_{\rm rel}}$, where $\mathbf{p_{\rm rel}}$ is the relative momentum between the particles. In a thermal medium this ratio is of the order of $T/(\sqrt{m_e \, T}) = \sqrt{T/m_e}$, where $T$ is the temperature of the plasma.  This means that for our temperatures of interest (around 1--10~keV), quadrupole processes will be suppressed by a factor $T/m_e \sim 10^{-2}$--$10^{-3}$. A further relative suppression derives from the ratio of ${\cal A}_2/{\cal A}_1$ discussed below.

Concerning the dipole term, from the commutation relation between the reduced Hamiltonian $H_0={\bf p}^2/2m+V({\bf r})$ with ${\bf r}$ one
finds that 
$\langle{\rm f}|{\bf p}|{\rm
  i}\rangle=i(E_{\rf}-E_{\ri})m\br_{\rf\ri}$ with 
$\br_{\rf\ri}=\langle{\rm f}|{\bf r}|{\rm i}\rangle$.
Therefore, we find for the dipole term
\begin{equation}\label{eq:Matrixelement-Photon-4}
  {\cal M}_{\rf\ri}=i{\cal A}_1\omega\,\be\cdot\br_{\rf\ri},
\end{equation}
where the emitted photon energy is $\omega=E_\ri-E_\rf$. In this way it is obvious that the matrix element is the same independently of the magnitude of the dipole moment ${\cal A}_1$ itself or how the two interaction partners contribute.

Beyond electromagnetism, we may imagine that the electron alone carries a ``leptonic charge'' $g_{\rm L}$ or the nucleons alone a baryonic one $g_{\rm B}$. In this case, the corresponding dipole moments are ${\cal A}_1^{\rm L}=g_{\rm L}/m_e$ or ${\cal A}_1^{\rm B}=g_{\rm B}/m_N$ with $m_N$ the nucleon mass. The ratio is ${\cal A}_1^{\rm B}/{\cal A}_1^{\rm L}=(g_{\rm B}/g_{\rm L})(m_{e}/m_{N})$. Therefore, apart from the obvious ratio of squared coupling constants, {\em the baryonic emission rate is suppressed by the squared-mass ratio} $(m_e/m_N)^2$. This insight is the main result of this discussion.

Quadrupole radiation is generically suppressed relative to dipole radiation as explained earlier. Moreover, in our exotic example, there is a factor ${\cal A}_2^{\rm B}/{\cal A}_2^{\rm L}=(g_{\rm B}/g_{\rm L})(m_{e}/m_{N})^2$ so that besides the squared ratio of coupling constants, the baryonic emission rate is suppressed by the ratio $(m_e/m_N)^4$. Therefore, the relative suppression is even larger and we may safely neglect quadrupole radiation in all cases of interest.

\subsection{Scalar radiation in analogy to photons}

We now pass to consider the scalar case, which is the focus of this work. For scalars $\phi$ interacting with electrons or nuclei, the potential
created by the radiation is simply $-g\,\phi$. Therefore, the matrix
element in the CM frame is
\begin{equation}\label{eq:Matrixelement-Scalar-1}
  {\cal M}_{\rf\ri}=-\Bigl\langle\rf\,\Big|g_1e^{-i\bq\cdot\br\,m_2/M}+g_2e^{i\bq\cdot\br\, m_1/M}\Big|\,i\Bigr\rangle.
\end{equation}
The first term 1 in the expansion of the exponentials does not lead to
radiation because the initial and final states are orthogonal. Expanding up to second order provides
\begin{equation}\label{eq:Matrixelement-Scalar-2}
  {\cal M}_{\rf\ri}=i\underbrace{m\biggl(\frac{g_1}{m_1}-\frac{g_2}{m_2}\biggr)}_{\textstyle{\cal A}_1}
  \langle\rf|\bq\cdot\br|\ri\rangle
  -\frac{1}{2}\underbrace{m^2\biggl(\frac{g_1}{m_1^2}+\frac{g_2}{m_2^2}\biggr)}_{\textstyle{\cal A}_2}
  \langle\rf|(\bq\cdot\br)^2|\ri\rangle.
\end{equation}
For electron-nucleus collisions and for a baryonic interaction, the
same hierarchy ${\cal A}_1\gg {\cal A}_2$ arises as in the baryonic
photon case discussed earlier. The dominant dipole term can be written
in the form 
\begin{equation}\label{eq:Matrixelement-Scalar-4}
  {\cal M}_{\rf\ri}=i{\cal A}_1\omega\,\hat{\bq}\cdot\br_{\rf\ri},
\end{equation}
where $\hat{\bq}$ is a unit vector in the direction of the emitted radiation and $\omega$ its frequency. This is precisely the same form as for photon emission Eq.~\eqref{eq:Matrixelement-Photon-4} with the replacement $\be\to\hat{\bq}$. The angular integration of the squared
matrix element leads to a factor $1/3$ in both cases, but in the emission of photons a factor of 2 appears for two polarization states. Otherwise the emission of a scalar or vector is the same.

Therefore, considering scalar leptonic vs.\ baryonic emission we conclude, in analogy to the vector case, that the baryonic dipole emission rate is relatively suppressed by the factor $(m_e/m_N)^2$ on top of the squared coupling-constant ratio.

Similar techniques have been used to relate low-energy emission processes of radiation with different spin parities. For example, the spectral axion emission from the Sun was found by similar scaling laws from the tabulated optical opacity \cite{Redondo:2013wwa}, and the same can be achieved for keV-range neutrino emission \cite{Vitagliano:2017odj}. Likewise, incomplete axion free-bound transition rates in the Sun or Earth were corrected using systematically the methods of non-relativistic quantum mechanics~\cite{Pospelov:2008jk}.

\subsection{Classical limit and Larmor formula}

The question of scalar radiation by the heavy partner in a binary collision has two aspects. One is to compare scalar with vector radiation and the other is the unusual kinematics, where the light partner provides the energy that can be radiated, whereas the heavy partner is the one doing the radiation.

Scalar bremsstrahlung in the classical limit was discussed, for example, by Ren and Weinberg \cite{Ren:1993bs} who showed that the emitted power by an accelerated scalar charge moving on a prescribed trajectory is half that of the corresponding electromagnetic case, a point that also follows from our quantum-mechanical discussion in the non-relativistic limit, while Ren and Weinberg used general kinematics. 

The electromagnetic power radiated by an accelerated charge in the non-relativistic limit is given by the Larmor formula~\cite{Larmor:1897}. For scalar emission it is then in natural units
\begin{equation}\label{eq:Larmor}
    P=\frac{\alpha_\phi}{3}\,|\dot\bv|^2.
\end{equation}
For two particles interacting by a central potential, the mutual force is opposite equal. Because ``force = mass $\times$ acceleration,'' the acceleration for the two partners is inversely proportional to their respective mass. If only one of them radiates, assuming we have either a leptonic charge or a baryonic one, the radiation power is inversely proportional to the squared mass of who is radiating. This is the same conclusion that we reached earlier in the quantum mechanical discussion. 

Based on the Larmor formula we can also estimate the energy-loss rate of a plasma.
In fact, let us consider a gas of non-relativistic electrons and ions, these latter with electric charge $Z \, e$ and mass $m_i$. Let us also focus on the emission from the ions only. These experience an acceleration simply due to Coulomb interaction with electrons
\begin{equation}
|\dot\bv_i| \sim \frac{Z e^2}{m_i \, b^2},
\end{equation}
where $b$ is the impact parameter. The power emitted during a single ``collision" is therefore
\begin{equation}
P \simeq \frac{\alpha_\phi}{3}\frac{Z^2 e^4}{m_i^2 b^4}.
\end{equation}
In a plasma, rather then considering single particles, we must consider clouds of particles with number densities $n_i$ and $n_e$, respectively for ions and electrons. 

If the relative velocity between the particles is $v_{\rm rel}$, then the total number of collisions per unit volume for the ions during an interaction interval $\Delta t$ will be $ n_e \, n_i \, v_{\rm rel} \Delta t \, b \, 2\pi \, db$. Given the impact parameter and the relative velocity, the typical time scale is simply set by $\Delta t \sim b/v_{\rm rel}$. The total emitted power is then
\begin{equation}
\frac{dP}{dV} \simeq \frac{\alpha_\phi}{3}\frac{Z^2 e^4}{m_i^2} 2\pi \, n_e \, n_i \, \int_{\rm b_{\rm min}}^{\rm b_{\rm max}} \frac{db}{b^2} \simeq \frac{\pi \alpha_\phi \, Z^2 e^4}{m_i^2} \frac{n_e n_i}{b_{\rm min}} \simeq \frac{\pi \alpha_\phi \, Z^2 e^4}{m_i^2} n_e n_i v_{\rm rel} m_e,
\end{equation}
where in the last step we considered $b_{\rm min} \sim 1/(m_e v_{\rm rel})$ as set by the uncertainty principle. (In this ``classical" argument we actually do need a vestige of quantum mechanics.) 

In a thermal plasma with temperature $T$ one has $v_{\rm rel} \sim \sqrt{T/m_e}$. Therefore we are left with the energy loss rate per unit volume
\begin{equation}\label{Eq:ClassicSpectrum}
Q_i^\phi \equiv \frac{dP}{dV} \simeq \frac{\pi \alpha_\phi \, Z^2 e^4}{m_i^2 \,} n_e n_i \sqrt{m_e T}.
\end{equation}
In the next section we will see that this simple scaling with the temperature and the particle masses is indeed obtained by a rigorous quantum field theory computation. 
 
\subsection{Summary}

We have studied radiation from two interacting non-relativistic particles, notably with very different masses such as electron-proton interaction. The main concern was to understand the modification between the usual situation when the light particle (the electron) carries the radiating charge (e.g.\ a leptonic charge) and when the heavy particle carries the radiating charge (e.g.\ a baryonic charge). Of course, for the usual electromagnetic case, it is the electron which mostly radiates, although the proton contributes subdominantly to the radiating dipole moment. In the relevant dipole approximation, we have found that the only modification is a factor $(m_e/m_p)^2$ that can be easily gleaned from the classical Larmor formula or from quantum-mechanical perturbation theory. This factor was unfortunately missed in recent discussions of stellar energy losses, leading to excessively restrictive bounds on the coupling constant of new baryophilic scalars~\cite{Dev:2020jkh, Balaji:2022noj}.

Still, it remains somewhat surprising that the factor $(m_e/m_p)^2$ is the {\em only\/} modification, applying to free-free, free-bound or bound-bound transitions. In the latter (bremsstrahlung), the spectrum of the emitted radiation is the same in both cases because the phase-space factors are the same, including Pauli blocking in a degenerate stellar medium. It is only when relativistic modifications come in, for us in the semi-relativistic plasma of a white dwarf, that this simple scaling receives corrections as we will see in the following Section.

Other unusual cases of bremsstrahlung, where the heavy partner radiates, were studied in the recent literature, including photon emission by nuclei that are hit by a small-mass dark-matter particle
\cite{Kouvaris:2016afs}. The crucial point was that the bremsstrahlung spectrum extends to the maximum available energy, i.e., the kinetic energy carried by the light particle. Notice that in this situation,
the center-of-mass frame of the colliding particles is nearly identical with the rest frame of the heavy one and it is at first surprising that the full kinetic energy stored in the light particle can be emitted by the heavy one. Another example is coherent neutrino scattering on nuclei, where the photon endpoint carries information of the neutrino mass because the bremsstrahlung photon can take up all the energy of the incoming neutrino~\cite{Millar:2018hkv}.

\section{Electron-proton bremsstrahlung}
\label{Sec:el_proton}

We now turn to a quantum-field theory calculation of scalar bremsstrahlung emission from a stellar plasma. In particular, we consider a novel CP-even scalar $\phi$ that interacts with protons and electrons according to 
\begin{equation}
\mathcal{L} \supset g_{p} \phi \,\bar{p}p \quad \mathrm{and} \quad  g_e \phi \,\bar{e}e.
\end{equation}
For the moment, we leave open if $g_e$ and $g_p$ are universal leptonic or baryonic ``charges'' that we used in the previous section, or if they are related to each other, for example, by a Higgs-portal interaction or if protons and neutrons carry different coupling constants.
We focus on the bremsstrahlung process
\begin{equation}
e(\bk_1)+p(\bp_1)\rightarrow e(\bk_2)+p(\bp_2)+\phi(\bq),
\end{equation}
where $p$ and $e$ are respectively protons and electrons in the star of interest, with their appropriate thermal distributions. We will find that the energy-loss rate per unit volume caused by proton or electron bremsstrahlung scale as $Q_p^\phi/Q_e^\phi=(m_e/m_p)^2$ as anticipated with our more elementary arguments in the previous Section, except for small corrections in a white dwarf, where the electrons are semi-relativistic.

\subsection{Emission rate for general electrons conditions}

The squared amplitude for the bremsstrahlung from electron-proton collisions, the latter non-relativistic, and quasi-massless scalars ($m_\phi \ll T$), is
\begin{equation}\label{Eq:GeneralAmplitude}
\sum_{\rm spins}|\mathcal{M}|^2=\underbrace{16\, m_p^2 \, e^4 [(E_1+E_2)^2-(\bk_1-\bk_2)^2)]}_{\text{common factor}} \underbrace{\frac{g_i^2 m_i^2[Q\cdot(M_{i,1}-M_{i,2})]^2}{(Q\cdot M_{i,1})^2(Q\cdot M_{i,2})^2(M_{j,1}-M_{j,2})^4}}_{\substack{\text{depends on the emitting particle $i=e$ or $p$} \\ \text{and the ``spectator'' particle $j = p$ or $e$ }}},
\end{equation}
where $Q=(\omega,\bq)$ is the four-momentum of the emitted scalar, $M_{i,1}$ ($M_{j,1}$) and $M_{i,2}$ ($M_{j,2}$) are respectively the initial and final four-momenta of the particle $i(j)=e$ or $p$ emitting (not emitting) the scalar, either the electrons with four momenta $K_{1,2} = (E_{1,2}, \bk_{1,2})$ or the protons with four momenta $P_{1,2}$. (An extension to scalars with larger masses is provided in Appendix~\ref{App:Massive}.)
One can see from Eq.~\eqref{Eq:GeneralAmplitude} that in the limit of non-relativistic electrons, the two amplitudes are the same up to a factor $g_i^2 m_i^2$. 

This expression diverges for $(M_{j,1}-M_{j,2})^2\to0$, but in a plasma, charged particles are subject screening effects. The most naive inclusion of this effect would be to assume an effective in-medium mass for photons, but there is no simple fundamental method for treating screening effects to a consistent order of perturbation theory. We will return to this subject later and in Appendix \ref{App:screening}, whereas for the moment we simply augment the squared matrix element with a screening factor ${\cal S}(\Delta M_j)$ that will be made more precise later in the context of specific assumptions about the medium.

Armed with the amplitude squared, we can compute the energy loss per unit volume due to the production of scalars from the species $i$ in a generic stellar plasma,
\begin{equation}
\label{eq:Qphi}
\begin{aligned}
Q^{\phi}_{i} ={}&\frac{1}{(2\pi)^{11}}\int \frac{d^3\bp_1}{2 m_p}\frac{d^3\bp_2}{2 m_p}\frac{d^3\bk_1}{2 E_1}\frac{d^3\bk_2}{2 E_2} \frac{d^3{\bf q}}{2 \omega} \omega\,\delta^4(P_1 + K_1 - P_2 - K_2 -Q)\\
&\qquad\qquad\qquad\qquad\qquad\qquad\qquad\qquad\times f_p(\bp_1)f_e(\bk_1)[1-f_e(\bk_2)] \sum_{\rm spins}|\mathcal{M}|^2 \\
={}&\frac{e^4g_i^2 m_i^2}{2(2\pi)^{11}}\int d^3\bp_1\frac{d^3\bk_1}{ E_1}\frac{d^3\bk_2}{E_2} d\omega \, \omega^2 d\Omega_\phi f_p(\bp_1)f_e(\bk_1)[1-f_e(\bk_2)]\,\delta(\omega-E_1+E_2)\\ 
&\kern5em{}\times \frac{[Q\cdot(M_{i,1}-M_{i,2})]^2\,[(E_1+E_2)^2-(\bk_1-\bk_2)^2]}{(Q\cdot M_{i,1})^2(Q\cdot M_{i,2})^2(M_{j,1}-M_{j,2})^4}\,{\cal S}\bigl(M_{j,1}-M_{j,2}\bigr),
\end{aligned}
\end{equation}
where $m_p$ is the proton mass, and we already integrated over the momentum of the final protons using the delta function for momentum conservation, which in the long-wavelength approximation reads $\bp_1 + \bk_1 \simeq \bp_2 + \bk_2$. Moreover, we already simplified the energy delta function. In fact, energy conservation imposes 
\begin{equation}\label{Eq:Energy_conservation}
  \omega = \frac{1}{2m_p}(\bp_1^2 - \bp_2^2) + E_1 - E_2 \simeq E_1 - E_2,
\end{equation}
where in the last step we neglected the proton kinetic energies. This approximation applies to bremsstrahlung from either protons or electrons, i.e., it is always the electron providing the emitted energy as stressed earlier in Section~\ref{Scalar_brem}.

At this point, one should explicitly write down the squared amplitudes, but it becomes difficult to treat the emission from electrons and protons on the same footing. Here we provide our final master formulae for the two cases separately, which can be obtained after some tedious algebra. For the proton, the energy-loss rate per unit volume 
becomes
\begin{multline}\label{Eq:MasterFormulaProtons}
Q^{\phi}_{p} =\frac{\alpha^2\alpha_p n_pm_e^4}{3\pi^{2}m_p^2}\int_1^\infty dy_1\int_1^{y_1} dy_2\,\frac{1}{1+\exp\left(\frac{m_ey_1-\mu}{T}\right)}\,\frac{1}{1+\exp\left(-\frac{m_ey_2-\mu}{T}\right)}\\[1ex] 
{}\times \int_{-1}^{+1}dx_{12}\,
{\cal S}\bigl[-2(1-y_1y_2+ x_{12}\,z_1 z_2)\bigr]
\frac{z_1 z_2(z_1^2+z_2^2-2\, x_{12} \,z_1 z_2)(1+y_1y_2+x_{12} \,z_1 z_2)}{(1-y_1y_2+ x_{12}\,z_1 z_2)^2},
\end{multline}
where $x_{12}\equiv \hat{\bk}_1\cdot\hat{\bk}_2$ is the cosine of the angle between the initial and final electrons, and we introduced the adimensional variables $y_i\equiv E_i/m_e$ and $z_i \equiv |\bk_i|/m_e=\sqrt{\vphantom{|}\smash{y_i^2-1}}$. Notice that the argument in the structure function is simply $(M_{j,1}-M_{j,2})^2/m_e^2$, i.e., the squared quadrimomenta exchange in the Coulomb propagator, normalized by $m_e^2$.
We also introduced the electromagnetic fine-structure constant $\alpha=e^2/4\pi$ and the analogous one for the scalar interaction, $\alpha_p \equiv g_p^2/4\pi$.

If the scalar is emitted from the electron, we find a similar, but more cumbersome, expression,
\begin{equation}\label{Eq:MasterFormulaElectrons}
\begin{split}
Q^{\phi}_{e}={}&\frac{\alpha^2\alpha_e n_pm_e^2}{\pi^3}\int_1^\infty dy_1\int_1^{y_1}dy_2\,\frac{1}{(1+\exp\left(-\frac{m_ey_2-\mu}{T}\right))}\frac{1}{1+\exp\left(\frac{m_ey_1-\mu}{T}\right)}\\[1.5ex]
\times \,& \,
\int_{-1}^{+1} dx_{12}{\cal S}(z_1^2+z_2^2-2\,z_1 z_2x_{12})\\[1.5ex]
\times\,& \, \int_{-1}^{+1} dx_1\int_0^{2\pi}d\phi\frac{z_1 z_2(1+y_1y_2+x_{12}\,z_1 z_2)}{(z_1^2+z_2^2-2\,z_1 z_2x_{12})^2}\, \frac{(y_1-y_2-x_1\,z_1 +x_2\,z_2)^2}{(y_1-x_1 \,z_1)^2(y_2-x_2z_2)^2}\,
\end{split}
\end{equation}
where $\alpha_e \equiv g_e^2/4\pi$, $x_{1}\equiv \hat{\bk}_1 \cdot\hat{\bq}$ is the angle between the emitted scalar and the incoming electron, and $x_{2} \equiv \hat{\bk}_2 \cdot\hat{\bq} =\cos\phi\sqrt{1-x_1^2}\sqrt{1-x_{12}^2}+x_1x_{12}$ is the angle between the scalar and the outogoing electron.

Our results are easily generalized to the case in which electrons scatter off non-relativistic ions with atomic number $Z$ and mass number $A$. Let us consider the more generic Lagrangian
\begin{equation}
\mathcal{L} \supset  g_{p} \phi \,\bar{p}p + g_{n} \phi \,\bar{n}n \quad \mathrm{and} \quad \, g_e \phi \,\bar{e}e
\end{equation}
and the scattering process 
\begin{equation}
e+(Z, A)\rightarrow e +(Z, A)+\phi,
\end{equation}
where $(Z, A)$ is an ion with atomic number $Z$ and mass number $A$. In this case Eq.~\eqref{Eq:MasterFormulaProtons} and Eq.~\eqref{Eq:MasterFormulaElectrons} are easily modified introducing the ``effective" couplings
\begin{subequations}
 \label{Eq:RescalingIons}
   \begin{eqnarray}
\alpha_p \rightarrow \alpha_p^{\rm eff} &\equiv&  \frac{Z^2}{4\pi A^2}\, \Bigl[g_p Z + g_n (A - Z)\Bigr]^2,\\[1ex]
\alpha_e \rightarrow \alpha_e^{\rm eff} &\equiv& \alpha_e Z^2,
\end{eqnarray}
\end{subequations}
where the factor $Z^2$ comes from the ion electric charge,  $[g_p Z + g_n (A - Z)]^2$ is a coherence factor to be included in the nucleophilic case because in the long wavelength approximation all the nucleons emit radiation coherently, and the factor $1/A^2$ takes into account the fact that now the entire ion needs to be accelerated to emit radiation. 

We stress that, as long as nucleons are non-relativistic, the results in this section are \textit{exact} and they can be computed for any electron chemical potential and temperature of interest. In the following we provide compact formulae which apply for different limiting cases for the electrons conditions.

\subsection{Non-relativistic, non-degenerate electrons}

Let us now consider the case in which electrons are non-relativistic and non-degenerate. The simple results derived here apply with good precision to the Sun and horizontal-branch stars. In such a weakly correlated plasma, Coulomb screening is well approximated by Debye screening, resulting in the static structure function
\begin{equation}\label{eq:S-Debye}
    S(\bk)=\frac{\bk^2}{\bk^2+k_{\rm s}^2},
\end{equation}
where $\bk=\bk_1-\bk_2$ and $k_{\rm s}$ is the screening wave number. The screening scale receives contributions from both free electrons and ions. In the non-degenerate limit they are
\begin{equation}\label{eq:ks}
k_\rs^2 = \underbrace{\frac{4 \pi \alpha \, n_e}{T}}_{\text{electrons}} + \underbrace{\frac{4 \pi \alpha}{T} \sum_j n_j Z_j^2}_{\text{ions}}, 
\end{equation}
where $n_e$ is the electron number density and $n_j$ the number densities of ions with electric charge $Z_j e$. A degenerate electron gas is much more ``stiff'' with regard to electric polarization and so they contribute much less to screening.

With this screening prescription,  the squared amplitude in the non-relativistic non-degenerate 
limit reduces to
\begin{equation}\label{Eq:AmpScalarNR}
\sum_{\rm spins}|\mathcal{M}|^2 = \frac{g_i^2}{m_i^2}\underbrace{\frac{64 \, e^4 m_e^2 m_p^2(\bm{\hat{\beta}_\phi}\cdot(\bk_1-\bk_2))^2}{(\bk_1-\bk_2)^2[(\bk_1-\bk_2)^2+k_\rs^2]\bq^2}}_{\text{common factor}},
\end{equation}
and the energy loss rate per unit volume is
\begin{equation}
Q^{\phi}_{i} = \frac{2g_i^2 \alpha^2 n_p}{3\, \pi^3 m_i^2} \int_0^{\infty}dk_1 k_1^2 \int_0^{k_1} dk_2 k_2^2 \int_{-1}^{+1} dx \, \frac{f_e(\bk_1)}{k_1^2 + k_2^2 - 2 k_1 k_2 x +k_\rs^2}.
\end{equation}
If electrons are non-relativistic, the occupation number is given by a Maxwell-Boltzmann distribution
\begin{equation}
    f_e(\bk_1) = \frac{n_e}{2}\Big(\frac{2\pi}{m_e T}\Big)^{3/2} e^{-k_1^2/2m_e T},
\end{equation}
and the energy-loss rate is
\begin{equation}
\label{eq:nonrelnondeg}
Q^{\phi}_{i} = \frac{2 \sqrt{2} g_i^2 \alpha^2 n_p n_e}{3 \,(\pi\, m_e T)^{3/2} m_i^2} \int_0^{\infty}dk_1 k_1^2 \int_0^{k_1} dk_2 k_2^2 \int_{-1}^{+1} dx \,\frac{e^{-k_1^2/2m_e T}}{k_1^2 + k_2^2 - 2 k_1 k_2 x +k_\rs^2}.
\end{equation}
We therefore find a compact expression for the scalar emission rate per unit volume,
\begin{mymathbox}
[ams gather, title= \quad \, Bremsstrahlung 
 production rate for non-relativistic and non-degenerate electrons, colframe=blue!20!black]
 \label{Eq:MasterFormulaNonRel}
Q^{\phi}_{i}= \frac{16 \sqrt{2} \, \alpha_i \alpha^2 n_p n_e \sqrt{m_e T}}{3 \, \sqrt{\pi} \, m_i^2}F_{\rm s}(k_\rs),
\end{mymathbox}
\noindent where the coefficient is
\begin{equation}
F_{\rm s}(k_\rs)=\int_0^{\infty}\!\!du_1\int_0^{u_1}\!du_2 \, u_1u_2e^{-u_1^2}\log\!\left[1+\frac{4u_1u_2}{(u_1-u_2)^2+u_\rs^2}\right],
\end{equation}
where $u_i=k_i/\sqrt{2m_eT}$ and $u_\rs=k_\rs/\sqrt{2m_eT}$. We can see that Eq.~\eqref{Eq:MasterFormulaNonRel} indeed agrees with the scaling of the classical result of Eq.~\eqref{Eq:ClassicSpectrum}. One can perform a similar computation for the vector case. We find the same result of the scalar, multiplied by a factor of $2$ coming from the sum over the polarizations, as expected from the quantum mechanics computation in Section~\ref{Scalar_brem}. In Appendix~\ref{Sec:QMCalculationBorn} we also sketch the analogous quantum-mechanical calculation of the emission rate at second-order in perturbation theory, also in Born approximation. This calculation proceeds along the lines of Section~\ref{Scalar_brem}, but assuming the initial and final states to be plane waves, with the Coulomb interaction included as a perturbation on the free-particle Hamiltonian.

\subsection{Degenerate electrons}

We now turn to the case of degenerate electrons, which will be relevant for scalar production in RG cores and WDs. Our calculations closely follow those in Ref.~\cite{Raffelt:1989zt}, where one of us computed the production rate of electrophilic pseudoscalars from bremsstrahlung. For completeness and comparison, we also include this case here. Assuming the electrons are degenerate we can take their momenta to be $k_{1,2} \sim k_{\rm F}$, where $k_{\rm F}$ is the Fermi momentum. Furthermore, the momentum transfer between electrons can be approximated as $|\bk_1 - \bk_2|^2 \simeq 2 \, k_{\rm F}^2 (1 - x_{12})$. With these approximations, Eqs.~\eqref{Eq:MasterFormulaProtons} and~\eqref{Eq:MasterFormulaElectrons} reduce to
\begin{mymathbox}[ams gather, title= \qquad \qquad \quad Bremsstrahlung 
 production rate for degenerate electrons, colframe=blue!20!black]
 \label{Eq:MasterFormulaDegenerate}
    Q_i=\frac{2\alpha^2 \alpha_{i} T^2n_p}{9}
    \times
    \begin{cases}\hfill\displaystyle
    \frac{m_e^2}{m_p^2}~F_p(\beta_{\rm F})&\hbox{baryophilic scalars,}
    \\[2ex]
    \hfill F_e(\beta_{\rm F})&\hbox{leptophilic scalars,}
    \\[1ex]
    \displaystyle
    \hfill\frac{\pi^2}{5}\Big(\frac{T}{m_e}\Big)^2~F_a(\beta_{\rm F})
    &\hbox{axion-electron,}
    \end{cases}
\end{mymathbox}
\noindent where in each case
\begin{equation}\label{eq:Fi-def}
    F_i(\beta_{\rm F})=\int_{-1}^{+1}\!dx_{12}\,\frac{\hat S(x_{12})}{1-x_{12}}\,G_i(x_{12},\beta_{\rm F}).
\end{equation}
Here $\hat S(x_{12})$ is a function that takes care of screening as a function of scattering angle and would be unity without screening. In the non-relativistic limit ($\beta_{\rm F}=0$), the integral kernels are $G_i=1$, whereas in general they are
\begin{subequations}
\begin{eqnarray}\label{eq:kernel-p}
    G_p(x_{12},\beta_{\rm F})&=&\frac{2-\beta_{\rm F}^2\,(1-x_{12})}{2\,(1-\beta_{\rm F}^2)}
    \\[1ex]
    \label{eq:kernel-e}
    G_e(x_{12},\beta_{\rm F})&=&\frac{3 \, (1-\beta_{\rm F}^2)}{16\pi}\int_{-1}^{+1}\!\!dx_1\int_{0}^{2 \pi}\!\!\!d\phi \,
    \frac{(x_1-x_2)^2\,\bigl[2-\beta_{\rm F}^2(1-x_{12})\bigr]}{(1-x_{12})(1-\beta_{\rm F}x_1)^2(1-\beta_{\rm F}x_2)^2},
    \\[1ex]
    \label{eq:kernel-a}
    G_a(x_{12},\beta_{\rm F})&=&\frac{3 \, (1-\beta_{\rm F}^2)}{16\pi}\int_{-1}^{+1}\!\!dx_1\int_{0}^{2 \pi}\!\!\!d\phi \,\frac{2(1-x_{12})-(x_1-x_2)^2}{(1-x_{12})(1-\beta_{\rm F}x_1)(1-\beta_{\rm F}x_2)},
\end{eqnarray}
\end{subequations}
where we recall that $x_{2}=\cos\phi\sqrt{1-x_1^2}\sqrt{1-x_{12}^2}+x_1x_{12}$. These expressions are even functions in $\beta_{\rm F}$ because the variables $x_1$ and $x_2$ vary homogeneously on the interval $-1$ to $+1$, so after integration, odd terms in $\beta_{\rm F}$ must disappear. The full analytic expressions are given in Appendix~\ref{App:kernels}, but they are too complicated to be illuminating. A low-order expansion of the $G$-functions in $\beta_{\rm F}$ is
\begin{subequations}
\begin{eqnarray}\label{eq:expansion-p}
    G_p(x_{12},\beta_{\rm F})&=&1+\beta_{\rm F}^2
    -\,\frac{\beta_{\rm F}^2}{2}\,\left(1-x_{12}\right)+{\cal O}\left(\beta_{\rm F}^4\right),
    \\[1ex]
    \label{eq:expansion-e}
    G_e(x_{12},\beta_{\rm F})&=&1+\beta_{\rm F}^2
    -\frac{9\beta_{\rm F}^2}{10}\left(1-x_{12}\right)+{\cal O}\left(\beta_{\rm F}^4\right),
    \\[1ex]
    \label{eq:expansion-a}
    G_a(x_{12},\beta_{\rm F})&=&1+\frac{\beta_{\rm F}^2}{5}
    -\,\frac{\beta_{\rm F}^2}{2}\,\left(1-x_{12}\right)+{\cal O}\left(\beta_{\rm F}^4\right).
\end{eqnarray}
\end{subequations}
For a RG near helium ignition or WDs with masses of around $0.6\,M_\odot$, a typical average density is $10^6~{\rm g}~{\rm cm}^{-3}$ and the composition is either of $^4$He or $^{12}$C and $^{16}$O, in all cases with $Y_e=Z/A=1/2$ electrons per baryon. In this case the Fermi momentum is $k_{\rm F}=409$~keV and the velocity at the Fermi surface is \smash{$\beta_{\rm F}=k_{\rm F}/(k_{\rm F}^2+m_e^2)^{1/2}=0.625$} and thus $\beta_{\rm F}^2=0.39$ is not very small so that relativistic corrections are not completely negligible.

The scaling of axion emission in Eq.~\eqref{Eq:MasterFormulaDegenerate} agrees with the general finding that in the non-relativistic limit, the bremsstrahlung axion emission rate is $\frac{1}{2}(\omega/m_e)^2$ that of the photon one \cite{Redondo:2013wwa}, whereas we found that the scalar emission rate is $\frac{1}{2}$ times that of photons. In other words, the non-relativistic scalar and pseudoscalar ones are the same up to a factor $(\omega/m_e)^2$ in the latter. The overall factor $(\pi^2/5)(T/m_e)^2\simeq2.0\,(T/m_e)^2$ in the integrated rate reflects that in bremsstrahlung $\langle\omega\rangle\simeq T$ is rather soft, but harder for axions than for scalars.

\subsection{Screening effects for red-giant conditions}

The actual emission rate strongly depends on how we deal with screening effects that are discussed in more detail in Appendix~\ref{App:screening}. The degenerate electrons are ``difficult to polarize'' and essentially form a neutralizing homogeneous background charge density in which the nuclei (or ions) are immersed. Neglecting screening by electrons and treating the nuclei as essentially static, screening is governed by the Debye limit of the static ion-ion structure function $S_{\rm i}(\bk)$ given in Eq.~\eqref{eq:S-Debye}. For a single species of nuclei with charge $Ze$, the ion-ion screening scale is given by $k_{\rm i}^2=4\pi\alpha Z^2 n_{\rm i}/T$. Expressing the screening function in our degenerate limit it terms of the scattering angle provides
\begin{equation}
    \hat{S}(x_{12})=\frac{1-x_{12}}{1-x_{12}+\kappa^2}
    \quad\hbox{where}\quad
    \kappa^2=\frac{k_{\rm i}^2}{2k_{\rm F}^2}
    =\left(\frac{4\rho}{9\pi\,m_u}\right)^{1/3}\frac{Z\alpha}{T}
=0.074\,\frac{Z\rho_6^{1/3}}{T_8},
\end{equation}
where $\rho_6=\rho/10^6~{\rm g}~{\rm cm}^{-3}$ and $T_8=T/10^8$~K. For a species with atomic weight $A$, the number density is $n_{\rm i}=\rho/A m_u$ with $m_u$ the atomic mass unit and we have assumed that $Z/A=1/2$. In a RG core before helium ignition, $T\simeq0.7\times10^8$~K 
\cite{Raffelt:1987yu} and for helium $Z=2$ so that $\kappa^2=0.21\ll1$. 

In the non-relativistic limit ($\beta_{\rm F}=0$), where all integral kernels are $G_i=1$, the dimensionless emission rates for all processes given by the Debye expression are
\begin{equation}
    F_{\rm D}=\int_{-1}^{+1}dx_{12}\,
    \frac{1-x_{12}}{1-x_{12}+\kappa^2}=\log\left(1+\frac{2}{\kappa^2}\right)\Big|_{\kappa^2=0.2}=2.40.
\end{equation}
On the other hand, using the integral kernels of Eqs.~\eqref{eq:kernel-p}--\eqref{eq:kernel-a} we find for $\beta_{\rm F}=0.6$ the values
\begin{equation}
    F_p(0.6)=1.38\, F_{\rm D},
    \qquad
    F_e(0.6)=1.24\,F_{\rm D},
    \qquad
    F_a(0.6)=0.94\,F_{\rm D}.
\end{equation}
As one might have guessed from the expansions Eqs.~\eqref{eq:expansion-p}--\eqref{eq:expansion-a}, the relativistic corrections are largest for the baryophilic scalars and smallest for axions, in the latter case slightly reducing the Debye result.

\subsection{Screening effects for white-dwarf conditions}

However, our main interest are WDs where the Debye screening prescription is no longer appropriate. The degree of correlation among the ions is measured by the plasma parameter 
$\Gamma=Z^2\alpha/a_{\rm i}T$, which is the ratio of the ion-ion Coulomb interaction energy over their thermal kinetic energy. Here $a_{\rm i}$ is the ion-sphere radius given by $n_{\rm i}^{-1}=(4\pi/3)\,a_{\rm i}^3$. Numerically $\Gamma$ evaluates to 
\begin{equation}\label{eq:plasma-parameter-main}
\Gamma=\frac{Z^2\alpha}{a_{\rm i}T}
=\frac{Z^2\alpha}{T}\left(\frac{4\pi\rho}{3Am_u}\right)^{1/3}\bigg|_{A=2Z}
=\frac{Z^{5/3}\alpha}{T}\left(\frac{2\pi \rho}{3m_u}\right)^{1/3}
=1.80\,\frac{Z^{5/3}\,\rho_6^{1/3}}{T_7},
\end{equation}
where we have used $A= 2 \, Z$. The plasma parameter is also connected to our parameter $\kappa^2$ through
\begin{equation}\label{eq:kappa-Gamma-main}
    \frac{\Gamma}{\kappa^2}=\left(\frac{3\pi^2}{2}\right)^{1/3}\!Z^{2/3}=2.46\,Z^{2/3}.
\end{equation}
Strong correlations begin for $\Gamma\gtrsim 1$, corresponding to $\kappa^2\gtrsim0.12$ for $^{12}$C, which we may call the liquid phase. For $\Gamma\gtrsim178$, the ions begin to crystallize in a lattice. 

For the luminosity function of low-mass WDs, we are primarily interested in the liquid phase. The static structure function must be determined numerically as discussed in more detail in Appendix~\ref{App:screening}. In all cases (axions and scalars) the $F_i$ functions in Eq.~\eqref{Eq:MasterFormulaDegenerate} can be written as
\begin{equation}\label{eq:FitFunction-New}
F_{\rm fit}(\rho,\Gamma)=A(\rho)\,\Gamma^{-0.37} + B(\rho)\,\Gamma^{+0.03},
\end{equation}
and the coefficient functions are found to be well fitted by the functional form
\begin{subequations}\label{Eq:FunctionalFormNewMain}
\begin{eqnarray}
A(\rho)&=& a_{0} + a_{1} x + \frac{a_{2}}{(8-x)} + \frac{a_{3}}{(8-x)^2},
\\[1ex]
B(\rho)&=& b_{0} + b_{1} x + \frac{b_{2}}{(8-x)} + \frac{b_{3}}{(8-x)^2},
\end{eqnarray}    
\end{subequations}
where $x=\log_{10}(\varrho)$ with $\rho$ in units of ${\rm g}/{\rm cm}^3$. The numerical coefficients differ for different atomic charge $Z$ and different bosons. Their values, and an extended discussion, can be found in Appendix.~\ref{Sec:FullNumeric}
(see in particular Table~\ref{tab:FitParametersNew}).

\section{Bounds from the white-dwarf luminosity function}
\label{sec:bounds}
\subsection{Introduction}

New low-mass particles can be systematically constrained by their emission from hot stellar plasmas, leading to observable consequences for the evolution of various well-observed stars or classes of stars. For scalar particles, the intriguing phenomena of resonant conversion from longitudinal plasmons was proposed some years ago, leading to very restrictive bounds $g_{\rm L}<0.7\times10^{-15}$ and $g_{\rm B}<1.1\times10^{-12}$ based on the brightness of the tip of the red giant (RG) branch~\cite{Hardy:2016kme}.\footnote{We observe that a small error has crept into the resonant emission rate Eq.~(2.31) of Ref.~\cite{Hardy:2016kme} where the middle factors should read $k_{\omega_{\rm p}}^3\omega_{\rm p}$ instead of 
$k_{\omega_{\rm p}}^2\omega_{\rm p}^2$. We thank E.~Hardy for confirming this point. However, it makes no difference for scalar bounds in the massless limit.} The difference between the leptonic and baryonic bounds actually represents the ratio of $g_{\rm L}/m_e$ and $4g_{\rm B}/m_{{}^4{\rm He}}$ that is now familiar from our bremsstrahlung argument.

Far more restrictive bounds were derived in Refs.~\cite{Dev:2020jkh,Balaji:2022noj} using 
{\em inter alia} bremsstrahlung emission in WDs. Unfortunately, for baryonic scalars they did not include the generic $m_e/m_p$ factor that we have argued in Sections~2 and~3. In addition, they did not use degeneracy effects correctly, another motivation to revisit the WD argument because it continues to provide one of the most restrictive limits even after these corrections.

WDs often provide very restrictive limits because, while they are about as hot as the Sun inside (around 1 keV) and have perhaps a half solar mass, they are only about the size of the Earth and therefore very dim because of their small surface, despite of being very hot (``white''). Therefore, volume particle emission competes only with a small photon surface luminosity. Moreover, they no longer burn nuclear fuel so that their evolution is a benign cooling process. Probably these points were made for the first time nearly 40 years ago in the PhD work of one of us \cite{Raffelt:1985nj} in the context of axion emission. Since that time, many authors have studied specifically axion emission from WDs, sometimes even observing a tentative excess cooling that can manifest itself in a drift of the oscillation frequency of variable WDs and in some cases this drift can be amazingly well measured. For more details and references to the original literature we refer to a recent review by some of the original authors \cite{Isern:2022vdx}.

In our argument, we will primarily use the WD luminosity function (WDLF) in the galactic disk \cite{Bertolami:2014wua,garcia2016white}, i.e., the distribution of WDs as a function of luminosity. WDs are the compact remnants of low-mass stars after undergoing the red-giant and asymptotic-giant phase, after which they tend to shed their envelope in the form of a beautiful planetary nebula and remain as a glowing ember. Assuming an approximately constant birth rate, the number of WDs in each brightness interval is a direct measure of the cooling speed that can be enhanced by scalar emission. Moreover, the slope of the luminosity function would be very different compared to the case when standard surface photon emission dominates. Because scalar emission scales with 2 fewer powers of $T$ compared with axions, the situation is now very different because axions change the shape of the luminosity function in much more subtle ways. We now venture to elaborate these general arguments in some detail.

\subsection{Rescaling of axion bounds}

However, before turning to a detailed analysis, we can get a rough idea of what to expect using the results obtained for axions. In fact, we have seen above that the functional forms of the emission rates are very similar, Eq.~\eqref{Eq:MasterFormulaDegenerate}, and therefore easy to rescale.

The most stringent axion bound of $g_{ae}<1.6\times10^{-13}$ at a nominal 95\% C.L.\ was derived from the brightness of the tip of the RG branch \cite{Capozzi:2020cbu}. With the scaling of Eq.~\eqref{Eq:MasterFormulaDegenerate} and using $F_{e}\simeq1$ and $F_{ae}\simeq1$ and $T=10^8~{\rm K}=8.6~{\rm keV}$, the RG axion bound translates to a bound on the scalar electron coupling of $g_e<4\times10^{-15}$, to be compared with a more stringent RG bound from resonant plasmon conversion of $g_e<0.7\times10^{-15}$ \cite{Hardy:2016kme}.

For axions, a comparable bound of $g_{ae}<2.8\times10^{-13}$ at a nominal 99\% C.L.\ derives from the WDLF \cite{MillerBertolami:2014rka}. From the scaling observed in the previous section, we can foretell the expected sensitivity for scalars. In fact, from Figs.~4 or~6 of Ref.~\cite{MillerBertolami:2014rka} one gleas that axion cooling gets constrained mainly by WDs with bolometric brightness $M_{\rm bol} \sim 7$--9, where
\begin{equation}
  M_{\rm bol}=4.74-2.5 \log_{10}(L/L_\odot).
\end{equation}
This range corresponds to internal $T \sim 2$--$3\,\rm keV$. Therefore, using again the $\beta_{\rm F}\to0$ limit, the bound on scalars would be $g_{e}< g_{ae} \times \sqrt{2}T/m_e \sim 10^{-15}$. This is only a rough estimate because scalar emission affects the WDLF in different ways compared to pseudoscalars. In particular, scalars are more important for colder and thus older WDs. In any case, this simple scaling suggests that the WDLF can provide a limit comparable to that from resonant plasmon conversion in RGs.

In both RGs and WDs we can also rescale the axion bounds to the scalar baryon case with the scaling factor $(m_e/m_p)^2$. For a scalar coupling to baryon number, the radiating ``charge'' of a nucleus is enhanced by a factor of its atomic number, but its mass receives the same factor, so indeed the scaling is $(m_e/m_p)^2$, independently of the chemical composition of the RGs (mostly helium) and WDs (mostly carbon and oxygen). Therefore, the estimated bremsstrahlung bounds from RGs would be
$g_{\rm B}\lesssim7\times10^{-12}$, to be compared with the more stringent
$g_{\rm B}<1.1\times10^{-12}$ based on resonant plasmon conversion \cite{Hardy:2016kme}.
Our naive scaled WD bound is $g_{\rm B}\lesssim0.6\times10^{-12}$, again suggesting that the WDLF can give stringent constraints.

\subsection{Energy-loss rates}

\begin{figure}
    \centering
    \includegraphics[scale=0.5]{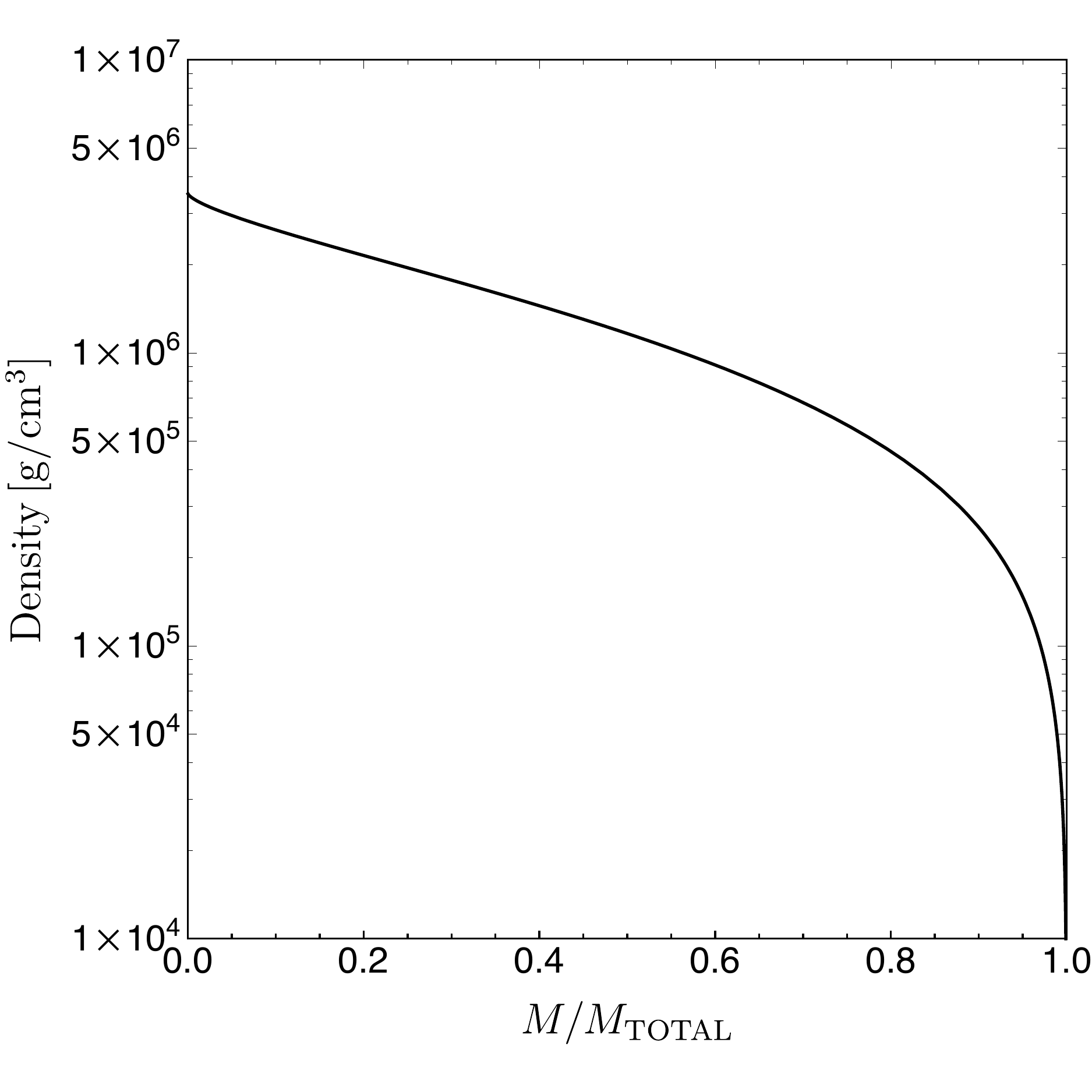}
    \caption{Density profile assumed for a typical $0.607\,M_\odot$ WD.}
    \label{fig:profile}
\end{figure}

Here we provide for clarity the explicit energy-loss rates per unit mass for the WD case $\epsilon^\phi_i \equiv Q^\phi_i / \rho$. We have checked that the degenerate approximation works extremely well and one can use directly Eq.~\eqref{Eq:MasterFormulaDegenerate}, instead of the more cumbersome general equations. We consider an equal mixture of carbon and oxygen. Assuming $g_n = g_p \equiv g_B$ for the baryonic case, one has
\begin{subequations}\label{Eq:RatesForWDs}
\begin{eqnarray}
    \epsilon^\phi_B&=&\frac{2\alpha^2 \alpha_B T^2}{9 \, {m_p}} \times \frac{m_e^2}{m_p^2} \times {\sum_j X_j\frac{Z_j^2}{A_j}}F_{p,j}(\rho,T),
    \\[1ex]
    \epsilon^\phi_e&=&\frac{2\alpha^2 \alpha_e T^2}{9 \, {m_p}} \times {\sum_j X_j\frac{Z_j^2}{A_j}}F_{e,j}(\rho,T),
\end{eqnarray}
\end{subequations}
where $X_j$ is the mass fraction of the element $j$, and the function $F_{i,j}$ are given for carbon and oxygen by the fitting formulae of Eq.~\eqref{eq:FitFunction-App} and we made explicit the dependence on temperature through Eq.~\eqref{eq:plasma-parameter}. The energy-loss rate per unit mass can be integrated over the entire stellar profile to yield the total scalar luminosity,
\begin{equation}
    L^{\phi}_i=\int_{M_{\rm WD}}\epsilon_i^\phi(\rho, T)\, dM.
\end{equation}
The WD core can be considered isothermal. The density profile can be obtained enforcing hydrostatic equilibrium and mass conservation (see e.g. Ref.~\cite{Shapiro:1983du,koester1990physics} and Section 3.5 of Ref.~\cite{hansen2012stellar}),
\begin{equation}
    \frac{dP(r)}{dr}=-\frac{GM(r)\rho(r)}{r^2}
    \quad\hbox{and}\quad
    \frac{dM(r)}{dr}=4\pi r^2\rho(r),
\end{equation}
where $M(r)$ is the enclosed mass, $P(r)$ is the pressure, and $\rho(r)$ is the density at radius $r$. The two boundary conditions are used to fix e.g.\ the pressure at the core boundary and the central density. Finally, the equation of state can be described assuming that electrons form an ideal Fermi gas with approximately zero temperature that prevents the star from its gravitational collapse. Defining the dimensionless ``relativity parameter'' $x\equiv k_{\rm F}/m_e$, one finds
\begin{equation}
    P=\frac{m_e^4}{24\pi^2} \left[x \sqrt{1+x^2}\left(2x^2-3\right)+3\log\left(x+\sqrt{1+x^2}\right)
    \right].
\end{equation}
In the zero-temperature limit, $k_{\rm F}^3=3\pi^2 Y_e \rho/m_u$ 
 with $Y_e=0.5$, and the system can be solved. We will make the crude assumption that all WDs have a central density $\rho_c=3.5\times 10^6\,\rm g/cm^3$, which corresponds to $M_{\rm WD}=0.607 \, M_\odot$, approximately the average mass of DA WDs~\cite{Tremblay:2009uv,Limoges:2010vm,Kleinman:2012nt}, that constitute the largest population of WDs. The inclusion of General Relativity and Coulomb corrections can be neglected at our level of accuracy. We show in Fig.~\ref{fig:profile} the profile in mass coordinates,  that one can compare with Fig.~1 of Ref.~\cite{Bischoff-Kim:2007yxc}, the density profile of a $0.602\, M_\odot$ WD. We conclude that our approximations should give a representative profile up to perhaps a few tens percent.

\subsection{WD cooling and luminosity function}
\label{subsec:bounds}

WDs have no nuclear energy sources and their evolution is basically a cooling process, based on the emission of photons and neutrinos, and potentially of new particles $X$. The number density of WDs in a given magnitude interval is (see e.g. Eq.~2.9 of Ref.~\cite{Raffelt:1996wa}, and Ref.~\cite{Shapiro:1983du})
\begin{equation}\label{eq:almostmestel}
\frac{dN}{dM_{\rm bol}} = B_3 \, 2.2 \times 10^{-4} \, \frac{10^{-4M_{\rm bol}/35}L_\odot}{78.7 \, L_\odot 10^{-2M_{\rm bol}/5}+L_\nu + L_X} \Big(\frac{M}{M_\odot}\Big)^{5/7} \Big(\sum_j \frac{X_j}{A_j}\Big) \, \rm pc^{-3} mag^{-1}, 
\end{equation}
where $B_3$ is the (constant) birthrate normalised to $10^{-3} \rm \, pc^{-3} Gyr^{-1}$. There are a number of assumptions needed to obtain Eq.~\eqref{eq:almostmestel}.   The relationship between the surface luminosity and the internal temperature is obtained assuming the Kramer's opacity, so that $L_\gamma=C_\gamma L_\odot T^{7/2}$, where $T$ is the temperature in the core and $C_\gamma$ is determined by fitting the data of the WDLF corresponding to photon-dominated cooling. Moreover, the thermal energy is considered to be stored in the nuclei. We neglect additional effects such as physical separation processes, convection, the contribution of electrons to the specific heat, and magnetic fields~\cite{Althaus:2010pi,Drewes:2021fjx}.

The cooling of hot WDs is dominated by neutrino emission through plasmon decay~\cite{Winget:2003xf}. However, once the WD is cool enough, plasmons get suppressed and cooling is dominated by photon emission from the surface. Neglecting for the time being also $L_X$ one finds
\begin{equation}
\frac{dN}{dM_{\rm bol}} = B_3 \, 2.9 \times 10^{-6} \, 10^{2\, M_{\rm bol}/7}\Big(\frac{M}{M_\odot}\Big)^{5/7} \Big(\sum_j \frac{X_j}{A_j}\Big) \, \rm pc^{-3} mag^{-1}.
\end{equation}
Assuming an equal mixture of carbon and oxygen, and taking $M = 0.6 \, M_{\odot}$ and $B_3 = 1$, we obtain Mestel's cooling law~\cite{mestel1952theory}
\begin{equation}
 \log_{10}\,(dN/dM_{\rm bol}) = \frac{2}{7} \, M_{\rm bol} - 6.84 +  \log_{10}(\rm B_3),
\end{equation}
which provides a very good fit to data for intermediate luminosity. In this region, the luminosity can be written as $L=C_\gamma L_\odot T^{7/2}$, with $C_\gamma=8.5\times 10^{-4}$~\cite{Giannotti:2015kwo}. One can therefore obtain the core temperature by inverting this relationship, which in principle is valid only as far as cooling is dominated by photon emission.

\begin{figure}[H]
    \centering
    \includegraphics[scale=0.5]{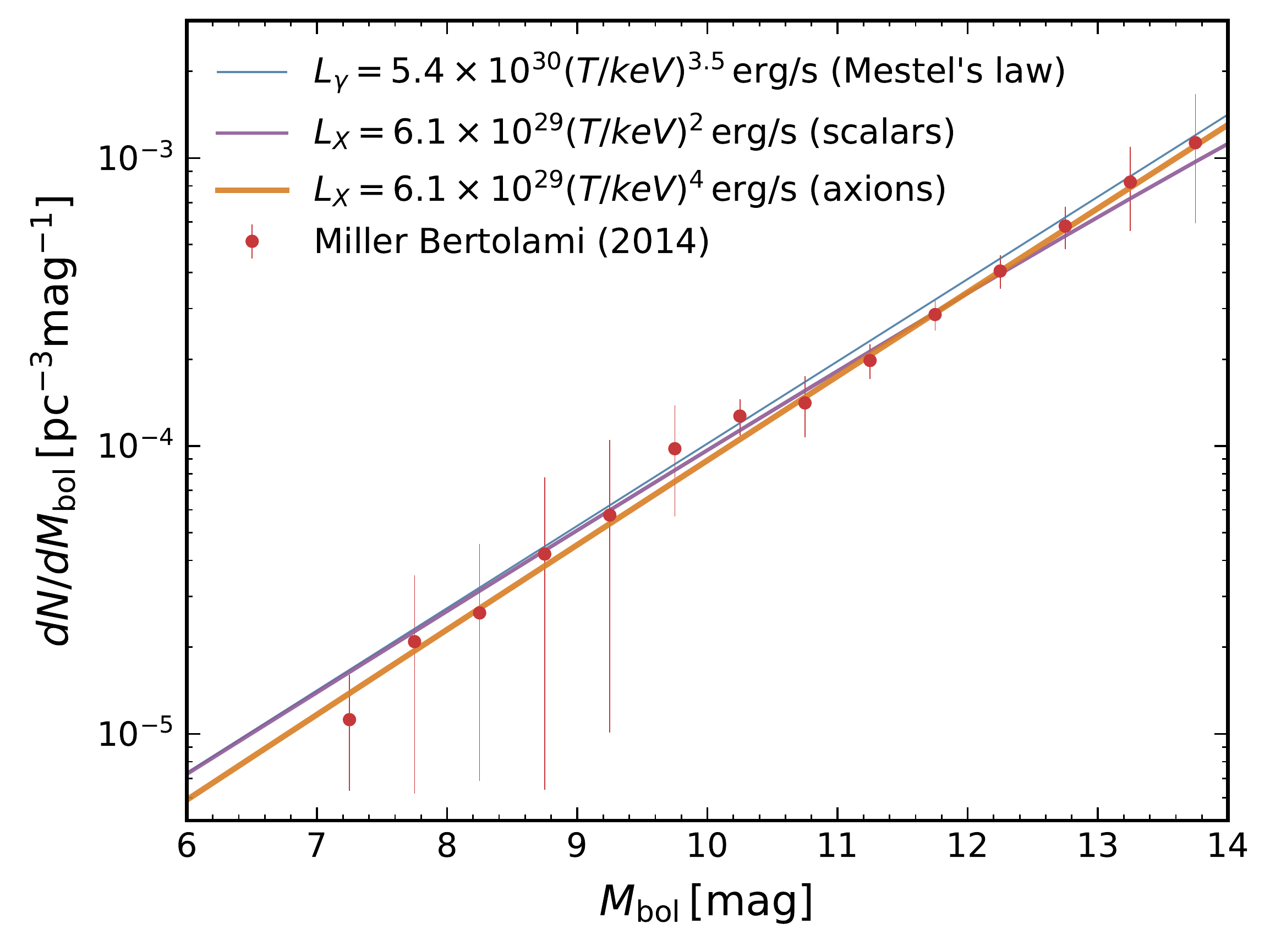}
    \caption{Comparison between the WDLF data (red dots) from Ref.~\cite{MillerBertolami:2014oki}, the simple Mestel's law (thin blue) and a WDLF in the presence of an extra cooling due to axions (orange) or scalars (purple). The scalar case departs from Mestel's law at large $M_{\rm bol}$, while the pseudoscalar case at small ones. For all curves the WD birth rate was fixed to $B_3 = 1$.}
    \label{fig:WDLFData}
\end{figure}

The emission of novel feebly interacting particles modifies Mestel's cooling law. As we assumed the WD core to be isothermal, the effect of scalars can be parametrized by the $T$ dependence of the energy-loss rate, and the coupling of scalars to electrons or protons. In Fig.~\ref{fig:WDLFData}, we show the WDLF data with $3\, \sigma$ error bars from Ref.~\cite{MillerBertolami:2014oki}, together with Mestel's law (thin blue), and two curves corresponding to scalar emission (thick purple) and axion emission (thick orange), parametrized as $L_X = L_{X,0} T_{\rm keV}^{\,n}$. Scalar and pseudoscalar emission rates can be parametrized respectively with $n = 2$ and $n = 4$. From Fig.~\ref{fig:WDLFData} we see that the effect of axions is particularly pronounced at smaller $M_{\rm bol}$ (hotter WDs), while scalars kick in at lower internal $T$ (larger $M_{\rm bol}$).

Figure~\ref{fig:WDLFData} gives already an idea of the maximum extra cooling allowed by data, depending on temperature dependence. In order to be more quantitative, however, we run a simple statistical test. We consider the theoretical (``th") expression for the WDLF in the case of scalar emission from the species ``$i$" (electrons or baryons)
\begin{equation}\label{eq:almostmestel-2}
\left(\frac{dN}{dM_{\rm bol}}\right)_{{\rm th}, i} = \textcolor{red}{B_3}  \, \frac{2.2 \times 10^{-4} \, 10^{-4M_{\rm bol}/35}L_\odot}{78.7 \, L_\odot 10^{-2M_{\rm bol}/5}+ \textcolor{red}{g_i}^2 L^\phi_i(g_i = 1, T)} \Big(\frac{M}{M_\odot}\Big)^{5/7} \Big(\sum_j \frac{X_j}{A_j}\Big) \, \rm pc^{-3} mag^{-1}, 
\end{equation}
where we highlighted in red the two free parameters of the model. Given a WD model and an array of measured $M_{\rm bol}$, which in turn determines a temperature array, the scalar cooling is entirely determined up to a $g_i^2$ rescaling. We then consider the experimental data from Ref.~\cite{MillerBertolami:2014oki}, shown in Fig.~\ref{fig:WDLFData}, with their associated errors bars $\sigma(M_{\rm bol})$ and build a two-parameters $\chi^2$ statistic
\begin{equation}\label{Eq:Chi2Scalar}
\chi^2(g_i, B_3) = \sum_{M_{\rm bol} = 7.75}^{12.75} \frac{\Bigl[\Big(\frac{dN}{dM_{\rm bol}}\Big)_{\rm th, i} - \Big(\frac{dN}{dM_{\rm bol}}\Big)_{\rm exp} \, \Bigr]^2}{\sigma(M_{\rm bol})^2},
\end{equation}
where we use only data in the bolometric magnitude range $7.75 \, < M_{\rm bol} < 12.75$. Taking this subset of data is justified for two reasons. On the one hand, at low magnitudes neutrino cooling cannot be neglected. On the other hand, for large magnitudes (very cold WDs) crystallization effects become relevant. In fact, for very old and cold WDs, the ions begin to freeze into a regular lattice structure~\cite{Isern:1997na}. Nevertheless, for bolometric luminosity $M_{\rm bol} < 12.25$, crystallization should not be relevant yet~\cite{garcia2016white} and therefore our scalar emission rate is precise. Of course, a truly self-consistent treatment should closely follow the procedure of Ref.~\cite{MillerBertolami:2014oki}, and one should evolve WDs models which include the extra cooling process of scalar emission ab-initio. We shall perform a dedicated study in a future work, nevertheless, the present procedure should provide the correct ballpark for the excluded values. 

\begin{figure}[htp]
    \centering
    \includegraphics[width = 0.65\textwidth]{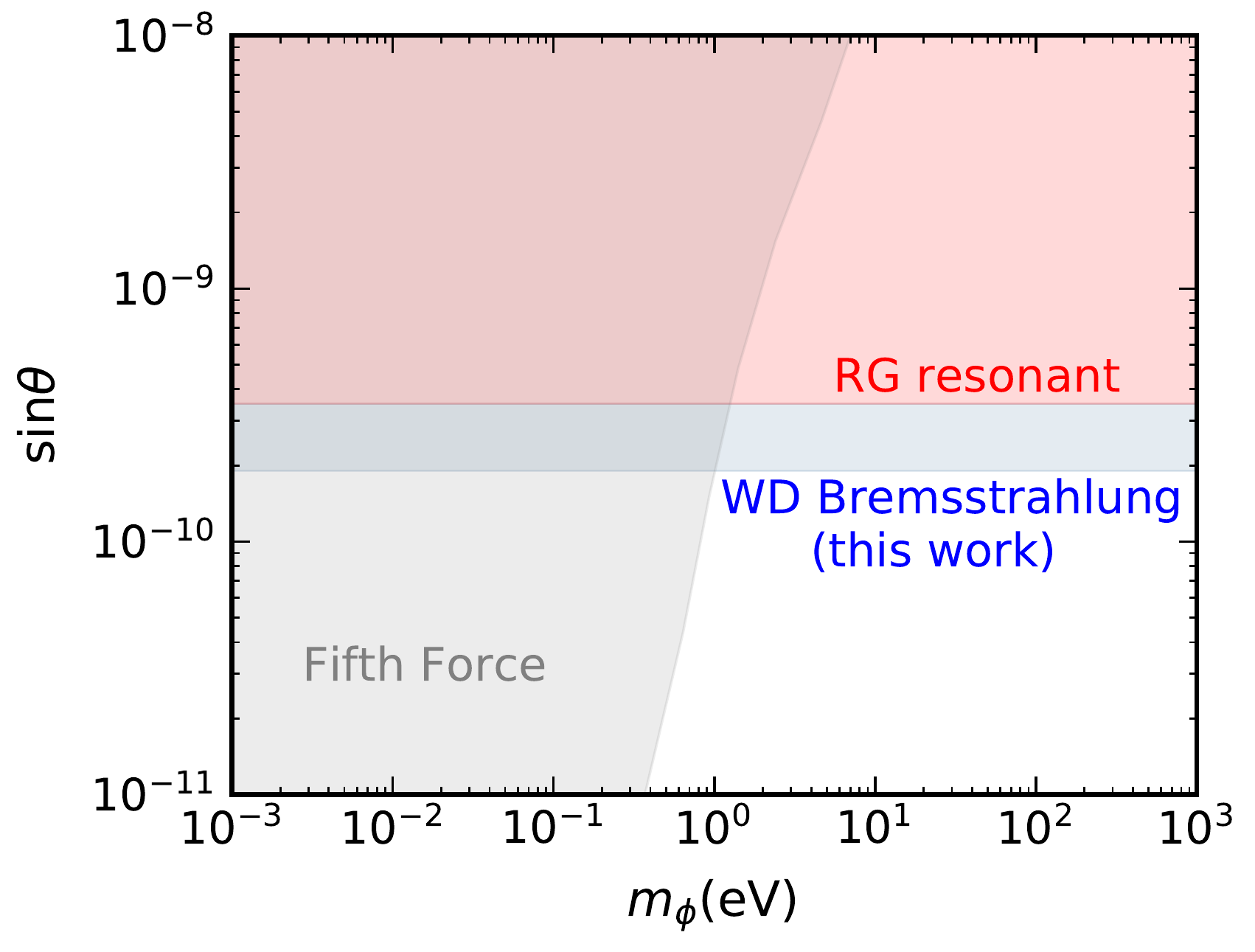}
    \caption{Constraints on the mixing angle $\sin\theta$ between a massive scalar $\phi$ and the SM Higgs. Fifth-force experiment bounds (gray region) are taken from Ref.~\cite{Flacke:2016szy}, while RG bounds from resonant conversion (red) from Ref.~\cite{Hardy:2016kme}.}
    \label{fig:HiggsMixingBounds}
\end{figure}

We therefore minimize Eq.~\eqref{Eq:Chi2Scalar} and find the exclusion limits for $g_i$. We assumed an equal mixture of carbon and oxygen. We checked that a one-zone model with constant density $\rho=1.3\times 10^6\,\rm g/cm^3$ and total mass $M=0.6\, M_\odot$ gives similar results. For a baryophilic scalar, the best fits are $B_3=1.02$ and $\alpha_B=1.15\times 10^{-26}$, with a reduced chi squared $\chi^2_{\rm red}=2.04$. For a leptophilic scalar, we find a best fit for $B_3=1.04$ and $\alpha_e=5.02\times 10^{-33}$, with a reduced chi squared $\chi^2_{\rm red}=2.03$. We find the nominal $95 \% \, \rm C.L.$ limits
\begin{subequations}\label{Eq:WDLimits}
\begin{eqnarray}
\alpha_B&\lesssim& 3.4 \times 10^{-26}, \\[1ex]
\alpha_e&\lesssim& 1.2 \times 10^{-32}.
\end{eqnarray}
\end{subequations}
For the corresponding Higgs portal case, we use $g_e=(m_e/v)\sin{\theta}$ and find
\begin{equation}
    \sin \theta \lesssim 1.9 \times 10^{-10}.
\end{equation}
In Fig.~\ref{fig:HiggsMixingBounds} we show this bound in the context of that from red giants and long-range force experiments.
Our WD bounds are still somewhat more restrictive than those derived from plasmon resonant conversion \cite{Hardy:2016kme}, but many orders of magnitudes weaker than those of Refs.~\cite{Balaji:2022noj,Dev:2020jkh} because of an incorrect emission rate. More precisely, their bound on $\sin\theta$ is 7~orders of magnitude more restrictive, i.e., a difference of 14 orders of magnitude in the emission rate.  The main sources of discrepancy in the emission rate are the factor $(m_e/m_p)^2 \sim 10^{-6}$ and their use of nondegenerate approximations for the WD environment, which leads to another missing factor $\sim (T/E_F)^2 \sim 10^{-6}$. These two factors only led to an overestimate of the scalar flux by roughly 12 orders of magnitude. Furthermore, the WD temperature and luminosity assumed for their one-zone model bound do not match at any point of Mestel's cooling law. Finally, their limit on $\sin\theta$ was derived from baryonic emission, while emission from electrons should prevail.

The bounds derived in this work are the strongest ones for $m_\phi \gtrsim \rm eV$, while for smaller masses fifth-force experiments prevail~\cite{Flacke:2016szy}. Nevertheless, a truly precise comparison between RG bounds and our new WD constraints is beyond the scope of this work and requires a dedicated effort. Neither here nor in Ref.~\cite{Hardy:2016kme} a robust statistical and astrophysical analysis has been undertaken. Furthermore, Ref.~\cite{Hardy:2016kme} assumed non-degenerate and non-relativistic electrons. This approximation allowed them to write the scalar self-energy in a very simple form, analogous to the photon self-energy. However, a RG core at helium ignition is degenerate and semi-relativistic like our WDs, only somewhat hotter. We expect this approximation to affect the RG bound at most by a factor of a few in coupling.

\section{Conclusions and outlook}\label{sec:conclusions}

Recent interest in the existence of putative scalar particles coupling to ordinary matter (electrons and nucleons) has prompted us to revisit stellar bounds, focusing on white dwarfs (WDs), which often offer competitive bounds on novel particles. The main production mechanism for both baryophilic and leptophilic scalars in WDs relies on electron-nucleus bremsstrahlung, motivating an explicit evaluation of the energy-loss rate due to this process. We have found the energy-loss rate for any plasma condition, and we have obtained compact expressions for the emission of scalars by a non-relativistic, non-degenerate plasma, as well as by a degenerate plasma for any degree of relativistic electron motion.

While the emission of scalars from electrons is conceptually very similar to the emission of electromagnetic radiation caused by the acceleration of the light charged particle in the collision, the emission of baryophilic scalars is less trivial, and has generated some confusion in the recent literature. We have shown that, while somewhat surprising, the only modification to obtain the baryonic emission rate from the electron emission rate (in turn related to the photon emission rate) is the inclusion of the factor $(m_e/m_p)^2$. This result applies to free-free, free-bound or bound-bound transitions and to any degree of electron degeneracy. This simple scaling applies only in the non-relativistic limit, whereas for the semi-relativistic conditions in WDs small corrections (tens of percent) arise.

Following earlier studies of axion emission, we have obtained novel bounds from the effect that the emission of scalars has on the WD luminosity function (WDLF) in the galactic disk. We have found that the recent evaluation of these bounds from WDs were overly stringent by several orders of magnitude, the difference arising from erroneous bremsstrahlung rates. Despite this reduction of sensitivity, the WDLF continues to provide one of the most restrictive limits, slightly more restrictive than the estimated bounds from resonant conversion of longitudinal plasmons in red giants at helium ignition existing in the literature.

Besides the specific constraints derived in our paper, we identify several directions for future work. For baryophilic scalars, it appears that the WDLF provides the most restrictive limits. To substantiate them, one should perform a self-consistent evolution of WD models, including the emission of scalars, that can have non-trivial effects on the WDLF that could not be captured by our simple treatment. To take advantage of the data from the dimmer end of the WDLF, both of common WDs as well as heavier WDs that show crystallization, one needs to compute the energy-loss rate of scalars in a strongly coupled plasma. We shall perform these computation in a future work.

For leptophilic scalars, the resonant conversion of longitudinal plasmons in the core of red giants at helium ignition looks like the most powerful argument. To substantiate these results, one needs to include degeneracy effects and semi-relativistic electrons in the plasmon conversion rate. It also would be interesting to evaluate directly the impact of this emission rate on the brightness of the tip of the red giant branch as it has been done for axions and neutrino dipole moments.

{\bf Note Added:} Shortly after our paper had appeared on arXiv, an independent study appeared that found similar conclusions with regard to scalar bremsstrahlung as well as to the relevance of white-dwarf cooling \cite{Yamamoto:2023zlu}.

\section*{Acknowledgements}

SB is supported by the Israel Academy of Sciences and Humanities \& Council for Higher Education Excellence Fellowship Program for International Postdoctoral Researchers. AC thanks Josef Pradler for useful conversations. This article is based upon work from COST Action COSMIC WISPers CA21106, supported by COST (European Cooperation in Science and Technology). GR\ acknowledges partial support by the German Research Foundation (DFG) through the Collaborative Research Centre ``Neutrinos and Dark Matter in Astro- and Particle Physics (NDM),'' Grant SFB-1258, and under Germany’s Excellence Strategy through the Cluster of Excellence ORIGINS EXC-2094-390783311. EV acknowledges support by the European Research Council (ERC) under the European Union's Horizon Europe research and innovation programme (grant agreement No. 101040019). Views and opinions expressed are however those of the author(s) only and do not necessarily reflect those of the European Union.

\appendix

\section{Details for the squared amplitude computation}
\label{App:DetailsAmplitude}

Given the recent confusion in the literature, we provide here a pedagogical derivation of the squared amplitude for the bremsstrahlung emission of a massless scalar. The process to be evaluated is $e(k_1) + N(p_1) \rightarrow e(k_2) + N(p_2) + \phi(q)$, and for simplicity we focus only on the nucleophilic case. The electrophilic computation proceeds in a similar fashion. 

Let us start with the relevant kinematic, where some subtleties are present. As explained in the main text, we assume the nucleons to be non-relativistic (an extremely good approximation in all the cases of interest), while we do not make any assumptions about the electrons. We can therefore define a small parameter, $\epsilon \propto \bp_i^2/m_N \ll 1$, and perform a perturbative expansion in this parameter for all the scalar products of interest. Let us also define the following quadri-momenta: $Q=(\omega,\bq)$ for the emitted scalar, $K_{1,2} = (E_{1,2}, \bk_{1,2})$ for the incoming and outgoing electron,  $P_{1,2} = (E^N_{1,2}, \bk_{1,2})$ for the nucleons. For the computation of the squared amplitude, few products will be needed and it is important to keep track of their order in the $\epsilon$ expansion. One has $K_i \cdot Q \propto \epsilon$, $P_i \cdot Q \propto \epsilon + \mathcal{O}(\epsilon^{3/2})$, $P_i \cdot P_j \propto 1 + \mathcal{O}(\epsilon)$, $P_i \cdot K_j \propto 1 + \mathcal{O}(\epsilon^{1/2})$, $K_i \cdot K_j \propto 1$. To obtain our main results it is enough to keep only the leading $\epsilon$ terms in these scalar products.

Let us now evaluate the relevant Feynman diagrams. Only two diagrams contribute to the simple process under consideration, one with the scalar attached to the outgoing nucleon and one with the scalar attached to the incoming nucleon. The sum of the two amplitudes reads
\begin{equation}
    \mathcal{M}_a + \mathcal{M}_b  = \frac{i g_p e^2}{(K_1 - K_2)^2}\mathcal{M}_{e,\mu}\mathcal{M}_N^\mu,
\end{equation}
\exclude{
\begin{equation}
\begin{split}
    \mathcal{M}_a + \mathcal{M}_b  = & \frac{i g_p e^2}{(K_1 - K_2)^2} \bar{u}_e(\bk_2)\gamma^{\mu}u_e(\bk_1)\bar{u}_N(\bp_2)\Big(\frac{(\slashed{P_2}+\slashed{Q}+m_N)\gamma_\mu}{2 P_2 \cdot Q} -  \frac{\gamma_\mu(\slashed{P_1}-\slashed{Q}+m_N)}{2 P_1 \cdot Q}\Big) u_N(\bp_1),\\
    = & \frac{i g_\phi e^2}{(K_1 - K_2)^2}\mathcal{M}_{e,\mu}\mathcal{M}_N^\mu,
\end{split}
\end{equation}
}%
where we have factorized the amplitude into a piece concerning electrons and one concerning the nucleons as follows:
\begin{equation}
\begin{split}
    \mathcal{M}_e^\mu&=\bar{u}_e(\bk_2)\gamma^{\mu}u_e(\bk_1)\ ,\\
    \mathcal{M}_N^\mu&=\bar{u}_N(\bp_2)\Big(\frac{(\slashed{P_2}+\slashed{Q}+m_N)\gamma_\mu}{2 P_2 \cdot Q} -  \frac{\gamma_\mu(\slashed{P_1}-\slashed{Q}+m_N)}{2 P_1 \cdot Q}\Big) u_N(\bp_1)\ .
\end{split}
\end{equation}
The amplitude squared and summed over spins is therefore
\begin{equation}\label{Eq:AmpSquared}
\sum_{\rm spin} |\mathcal{M}_a + \mathcal{M}_b|^2 =\frac{g_p^2 e^4}{(K_1 - K_2)^4} \left(\sum_{\rm spin}\mathcal{M}_e^\mu\mathcal{M}_e^{*\nu}\right)\left(\sum_{\rm spin}\mathcal{M}_{N,\mu}\mathcal{M}_{N,\nu}^*\right).
\end{equation}
The electron part is easily evaluated as
\begin{equation}
\sum_{\rm spin}\mathcal{M}_e^\mu\mathcal{M}_e^{*\nu}= 4 (K_1^\mu K_2^\nu + K_1^\nu K_2^\mu) +4 \, \, g^{\mu\nu}\Big(\bk_1 \cdot \bk_2 + m_e^2 - E_1 E_2\Big);   
\end{equation}
the expression for the nucleon part is more cumbersome, but when contracted with the electron piece -- using the scalar products defined above and \textit{keeping the lowest order in the $\epsilon$-expansion} -- we obtain Eq.~\eqref{Eq:GeneralAmplitude}.

\section{Master formula for bosons with mass}
\label{App:Massive}

In this appendix we generalize our results to the case of a scalar with mass $m_\phi$. If the radiated scalar is massless, then its energy is of the  order of $T$, and therefore kinematically small compared to the masses and momenta of the other particles. If the scalar mass was much larger than $T$, then this assumption need not be true, but of course the emission would be exponentially suppressed. Therefore we still assume that the energy and momentum of the scalar are small compared to the other energies. Under this assumption, we find that the energy loss rate in Eq.~\eqref{eq:Qphi} remains unchanged and therefore is
\begin{equation}
\begin{aligned}
Q^{\phi}_{i}(m_\phi) ={}&\frac{e^4g_i^2 m_i^2}{2(2\pi)^{11}}\int d^3\bp_1\frac{d^3\bk_1}{ E_1}\frac{d^3\bk_2}{E_2} d|\bq| |\bq|^2 \,  d\Omega_\phi f_p(\bp_1)f_e(\bk_1)[1-f_e(\bk_2)]\,\delta(\omega-E_1+E_2)\\ &
\times \frac{[Q\cdot(M_{i,1}-M_{i,2})]^2\,[(E_1+E_2)^2-(\bk_1-\bk_2)^2]}{(Q\cdot M_{i,1})^2(Q\cdot M_{i,2})^2(M_{j,1}-M_{j,2})^4}S(M_{j,1}-M_{j,1}).
\end{aligned}
\end{equation}
We now parameterize the scalar 4-momentum as $Q=\omega \, (1,\bm{\beta}_\phi)$, with $\bm{\beta}_\phi$  the scalar speed, so that the master formula Eq.~\eqref{Eq:MasterFormulaProtons} for the emission of baryophilic scalars thus becomes
\begin{eqnarray}\label{Eq:MasterFormulaNucleonMassive}
Q^{\phi}_{p}(m_\phi)&=&\frac{\alpha^2\alpha_p n_pm_e^4}{3\pi^{2}m_p^2}\int_{1\textcolor{red}{+\frac{m_\phi}{m_e}}}^\infty dy_1\int_{1}^{y_1\textcolor{red}{-\frac{m_\phi}{m_e}}} dy_2 
\,\frac{1}{\left[1+\exp\left(\frac{m_ey_1-\mu}{T}\right)\right]
\left[1+\exp\left(-\frac{m_ey_2-\mu}{T}\right)\right]}
\nonumber\\[2ex]
    &&\times\int_{-1}^{+1}\!\! dx_{12}{\cal S}(1-y_1y_2+x_{12}z_1 z_2)
    \frac{\textcolor{red}{\beta_\phi^3}z_1 z_2 (z_1^2+z_2^2-2x_{12}\, z_1 z_2)(1+y_1y_2+x_{12} \, z_1 z_2)}{(1-y_1y_2+x_{12}\, z_1 z_2)^2}
\nonumber\\
\end{eqnarray}
where we recall that $z_i = \sqrt{\vphantom{|}\smash{y_i^2-1}}$. Likewise, the master formula Eq.~\eqref{Eq:MasterFormulaElectrons} for the emission of leptophilic scalars is of similar form
\begin{eqnarray}\label{Eq:MasterFormulaElectronsMassive}
Q^{\phi}_{e}(m_\phi)&=&\frac{\alpha^2\alpha_p n_pm_e^2}{\pi^{3}}\int_{1\textcolor{red}{+\frac{m_\phi}{m_e}}}^\infty dy_1\int_{1}^{y_1\textcolor{red}{-\frac{m_\phi}{m_e}}} dy_2 
\,\frac{1}{\left[1+\exp\left(\frac{m_ey_1-\mu}{T}\right)\right]
\left[1+\exp\left(-\frac{m_ey_2-\mu}{T}\right)\right]}
\nonumber\\[1ex]
   &&\times\int_{-1}^{+1}\!\! dx_{12}{\cal S}(z_1^2+z_2^2-2\,z_1 z_2x_{12})
   \nonumber\\[1ex]
    &&\times\int_{-1}^{+1}\!\! dx_1\int_0^{2\pi}\!\!d\phi\,
  \frac{\textcolor{red}{\beta_\phi}z_1 z_2(1+y_1y_2+x_{12}\,z_1 z_2)(y_1-y_2-\textcolor{red}{\beta_\phi} x_1\,z_1 +\textcolor{red}{\beta_\phi} x_2\,z_2)^2}{(z_1^2+z_2^2-2\,z_1 z_2x_{12})^2(y_1- \textcolor{red}{\beta_\phi} x_1 \,z_1)^2(y_2-\textcolor{red}{\beta_\phi} x_2z_2)^2}.
  \nonumber\\
\end{eqnarray}
where in red we highlighted the difference as compared to the massless case. In terms of the variables $E_i \equiv y_i \, m_e$ we find for the scalar velocity
\begin{equation}
\beta_\phi = \frac{\sqrt{(y_1 - y_2)^2-m_\phi^2/m_e^2}}{y_1 - y_2}.
\end{equation}
These differences can be understood as follows. The integration lower limit on $y_1$ comes from the fact that $E_1=\omega+E_2\geq m_\phi+m_e$, while the upper limit on $y_2$ from $E_2=E_1-\omega\leq E_1-m_\phi$. Finally, from the parametrization of $Q$ we get the scaling:
\begin{equation}
    d|\bq| |\bq|^2\frac{[Q\cdot(M_{i,1}-M_{i,2})]^2}{(Q\cdot M_{i,1})^2(Q\cdot M_{i,2})^2}\sim d\omega \beta_\phi\mathcal{F}(\beta_\phi)
\end{equation}
where in the non-relativistic limit $\mathcal{F}(\beta_\phi)\propto\beta_\phi^2$, as can be seen neglecting the $y_1-y_2$ term in the last factor of Eq.~\eqref{Eq:MasterFormulaElectronsMassive}, thus explaining the extra factor of $\beta_\phi^2$ in Eq.~\eqref{Eq:MasterFormulaNucleonMassive}.

\begin{figure}[H]
\centering
\includegraphics[width=0.65\textwidth]{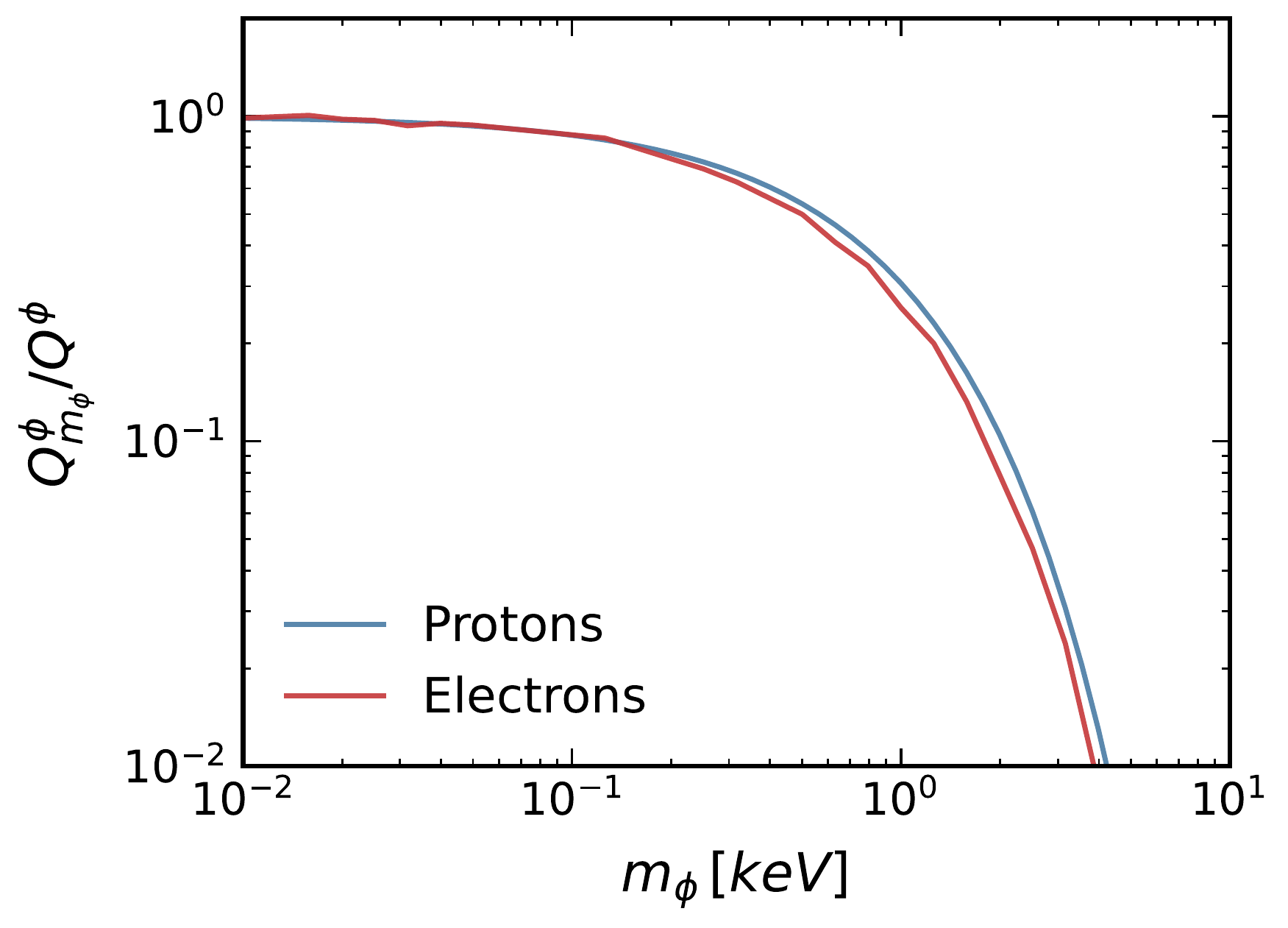}
\caption{Ratio of the emission rates for massive and massless scalars, as a function of the mass, assuming a density $\rho = 10^6 \rm \, g/cm^3$, temperature  $T = 1 \,\rm keV$ and screening scale $k_\rs = 1200 \, \rm keV$.
}  
\label{fig:Massive}
\end{figure}

In Fig.~\ref{fig:Massive} we show the ratio of the emission rates for the massive and massless cases as a function of the scalar $m_\phi$. For this plot we fixed the density of the medium to be $\rho = 10^6 \rm \, g/cm^3$, the temperature to $T = 1 \, \rm keV$ and the screening scale to $k_\rs = 1200 \, \rm keV$, which are typical values for the inner parts of WDs. It is evident that the production rate gets heavily suppressed as soon as $m_\phi \gtrsim T$. This is also why in the main text we limit our WDLF analysis up to masses $m_\phi \sim \rm keV$, given that $T \sim \rm keV$ is the typical temperature in the WD core.

\section{Explicit integral kernels}
\label{App:kernels}

The integral kernels Eqs.~\eqref{eq:kernel-e} and \eqref{eq:kernel-a} can be worked out explicitly. Setting  $x=x_{12}$ and $\beta=\beta_{\rm F}$ for compactness of typography, they are
\begin{eqnarray}
    G_e(x,\beta)&=&-\frac{3}{2}+\frac{3}{\beta^2 (1-x)}
    \\[1ex]
    &-&\frac{3 \left(1-\beta^2\right) \left[2-\beta^2 (1-x)\right]}
    {4 \beta^3 (1-x)^{5/2}\left[2-\beta^2 (1+x)\right]^{3/2}}
        \Biggl\{(2-x) {\rm ArcTanh}\left[\frac{\beta \sqrt{(1-x) \left[2-\beta^2
   (1+x)\right]}}{1-\beta^2 x}\right]
   \nonumber\\[1ex]
   \nonumber
   &&{}+\left[1-\beta^2 \left(1-x^2\right)-\frac{3 x}{2}\right]
   {\rm ArcTanh}\left[\frac{2 \beta \left(1-\beta^2 x\right) \sqrt{(1-x) \left[2-\beta^2
   (1+x)\right]}}{1+2 \beta^2 (1-2 x)-\beta^4 \left(1-2 x^2\right)}\right]\Biggr\}   
\end{eqnarray}
and
\begin{equation}
    G_a(x,\beta)=\frac{3}{2}\left(1-\beta^2\right)
    \left[\frac{\beta-{\rm ArcTanh}(\beta)}{\beta^3}
    +\frac{{\rm ArcTanh}\left[\frac{\beta\sqrt{(1-x) \left[2-\beta^2
   (1+x)\right]}}{1-\beta^2 x}\right]}
    {\beta\sqrt{(1-x) \left[2-\beta^2
   (1+x)\right]}}
    \right].
\end{equation}
We recall that
\begin{equation}
    {\rm ArcTanh}(y)=\frac{1}{2}\log\left(\frac{1+y}{1-y}\right)
\end{equation}
is an odd function of its argument. The use of these explicit kernels makes the numerical evaluation of the emission rates much faster.

\section{Bremsstrahlung in quantum mechanics using the Born approximation}\label{Sec:QMCalculationBorn}

We now connect the quantum-field theoretical calculation of the emission rate in Section~\ref{Sec:el_proton} (that uses the Born approximation) with the general quantum-mechanical argument about the mass scaling in Section~\ref{Scalar_brem}. To this end, we here sketch the quantum-mechanical calculation of the emission rate also in Born approximation. In Section~\ref{Scalar_brem} we assumed the initial and final wave functions of the interacting particles to be exact solutions of the interacting system before the interaction with scalars was included. So these could have been atomic wave functions or, in the free-free case, scattering states involving Coulomb wave functions. Now, on the other hand, we assume the initial and final states to be plane waves, whereas the Coulomb interaction itself is included as a perturbation on the free-particle Hamiltonian, implying that we need to go to second-order perturbation theory. Analogous computations for photon emission are found in Ref.~\cite{Gould:1981an} for example.

Specifically we consider electron-proton collisions and assume that the new scalar $\phi$ only couples to protons with a Yukawa strength $g_p$. The emission is therefore described by the total Hamiltonian
\begin{equation}
    H=\frac{\mathbf{k}_e^2}{2 m_e}  + \frac{\mathbf{p}_p^2}{2 m_p} + V_{\rm C}+V_{\phi}
\end{equation}
where the Coulomb and scalar-field potentials are
\begin{equation}
V_{\rm C}=-\frac{ \, \alpha}{|\mathbf{r}_e - \mathbf{r}_p|}
\quad{\rm and}\quad
V_{\phi} = - g_p \phi.
\end{equation}
We use rationalized units, where $\alpha=e^2/4\pi$. The amplitude for the process
$e(\bk_1)+p(\bp_1)\rightarrow e(\bk_2)+p(\bp_2)+\phi(\bq)$ arises at second order in the Born approximation and reads
\begin{equation}\label{eq:ampl}
    \mathcal{M}_{\rf\ri}=
 \mathcal{M}_{{\rm C},\phi}+\mathcal{M}_{\phi,{\rm C}}=
\sum_{a}\bra{\rf} V_{\rm C} \ket a\frac{1}{E_a-E_\ri}\bra{a} V_{\phi} \ket{\ri}+(V_{\rm C}\leftrightarrow V_\phi),
\end{equation}
where the sum is over intermediate states $\ket{a}$ with energy $E_a$. As usual, the sum includes on matrix element for the radiation emitted before and one after the Coulomb interaction.

We will now provide the different matrix elements without writing, for simplicity, the delta functions that enforce momentum and energy conservation. They will be reintroduced in the final result. The matrix elements of $V_{\rm C}$ are given by the Fourier transform of the Coulomb potential and are
\begin{equation}
    \bra{\rf} V_{\rm C} \ket{a}=\bra{a} V_{\rm C} \ket{\ri}=\frac{e^2}{|\mathbf{k}_1-\mathbf{k}_2|^2},
\end{equation}
because the exchanged momentum is always $\mathbf{k}_1-\mathbf{k}_2$ in the long-wavelength approximation where the momentum $\bq$ carried by the emitted radiation is ignored. Up to $\mathcal{O}(\omega)$ corrections, following the computations in Section \ref{Scalar_brem}, the matrix elements of $V_\phi$ are
\begin{equation}
    \bra{a} V_{\phi} \ket{\ri}=-\bra{\rf} V_{\phi} \ket{a}=\frac{ig_p}{\sqrt{2\omega}}\frac{\bm{\hat{\beta}}_\phi\cdot(\mathbf{k}_1-\mathbf{k}_2)}{m_p},
\end{equation}
where $\bm{\hat{\beta}}_\phi$ is a unit vector in the direction of motion of the emitted radiation.
We now compute the propagator $(E_\ri-E_a)^{-1}$ by enforcing momentum conservation. When the Coulomb scattering occurs \textit{after} the emission of the scalar as in $\mathcal{M}_{{\rm C},\phi}$, the scalar must be included in the intermediate state energy $E_a$, so that
\begin{equation}
\label{eq:prop_i}
  (E_\ri-E_a)\big|_{\mathcal{M}_{{\rm C},\phi}}=\frac{\bp_{1}^2-\bp_{a}^2}{2m_p}-\omega \approx\frac{\bq\cdot\bp_{1}}{m_p}-\omega.
\end{equation}
On the contrary, in $\mathcal{M}_{\phi,{\rm C}}$, the scattering occurs before emission, and the propagator is
\begin{equation}
\label{eq:prop_f}
   (E_\ri-E_a)\big|_{\mathcal{M}_{\phi,{\rm C}}}=\frac{\bp_{1}^2-\bp_{a}^2}{2 m_p}+\frac{\bk_{1}^2-\bk_{a}^2}{2m_e}\approx \omega -\frac{\bq\cdot\bp_{2}}{m_p}.
\end{equation}
Putting everything together, we find
\begin{equation}
    \mathcal{M}_{\rf\ri}= \frac{i g_p}{\sqrt{2\omega}}\,
    \frac{e^2}{|\mathbf{k}_1-\mathbf{k}_2|^2}\,
    \frac{2\bm{\hat{\beta}_\phi}\cdot(\mathbf{k}_1-\mathbf{k}_2)}{\omega \,m_p}.
\end{equation} 
Using \textit{non-relativistic phase space factors}, the energy loss rate is then given by:
\begin{eqnarray}
Q^\phi_p&=&\int\frac{d^3\bp_1}{(2\pi)^3}\frac{d^3\bp_2}{(2\pi)^3}\frac{d^3\bk_1}{(2\pi)^3}\frac{d^3\bk_2}{(2\pi)^3}\frac{d^3\bq}{(2\pi)^3}(2\pi)\delta(\omega-E_1+E_2)\, (2\pi)^3 \delta^3(\bp_1+\bk_1-\bp_2-\bk_2)\nonumber\\[1ex]
&&{}\times\omega  f_p(\bp_1)f_e(\bk_1)|\mathcal{M}_{\rf\ri}|^2
\nonumber\\[1ex]
&=&\int\frac{d^3\bp_1}{(2\pi)^3}\frac{d^3\bp_2}{(2\pi)^3}\frac{d^3\bk_1}{(2\pi)^3}\frac{d^3\bk_2}{(2\pi)^3}\frac{d^3\bq}{(2\pi)^3}(2\pi)\delta(\omega-E_1+E_2)\,(2\pi)^3 \delta^3(\bp_1+\bk_1-\bp_2-\bk_2)
\nonumber\\[1ex]
&&{}\times f_p(\bp_1)f_e(\bk_1)\omega \frac{g_p^2e^4}{2\omega|\mathbf{k}_1-\mathbf{k}_2|^4 }\frac{4(\bm{\hat{\beta}}_\phi\cdot(\mathbf{k}_1-\mathbf{k}_2))^2}{m_p^2 \, \omega^2}
\end{eqnarray}
which exactly matches the limit of non-relativistic electrons of Eq.~\eqref{eq:Qphi} except for the screening correction in the Coulomb propagator. 

\section{Screening prescription in the degenerate limit}
\label{App:screening}

\subsection{General formulation}

Particle emission from a medium is only approximately represented by individual processes among particles that interact as if they were in a vacuum. Even for simple examples such as Thomson scattering of photons on electrons in the Sun, one needs to include correlations to go beyond a rough estimate. Such correlations arise from the Pauli exclusion principle (an electron is less likely than average in the same location as another electron), but also from their Coulomb repulsion \cite{1987ApJ...316L..95B}, an effect that is easily overlooked in this context. For the bremsstrahlung processes discussed in Section~\ref{Sec:el_proton}, correlation effects are more dramatic because without them, the rate would diverge because of the infinite-range Coulomb interaction. On the other hand, it is clear that in an electrically neutral medium, the forward-scattering rate must vanish instead of diverge. For axion (pseudoscalar) emission, this question was explicitly addressed in Refs.~\cite{Itoh:1983, Nakagawa:1987pga, Nakagawa:1988rhp, Ichimaru:1987PhR, Raffelt:1989zt, Raffelt:1996wa} for the environments relevant in a RG core near helium ignition or in WDs, which are both electron-degenerate environments with $10^6\,{\rm g}\,{\rm cm}^{-3}$ range densities and temperatures in the $10^{6.5}$--$10^8$~K range, corresponding to $T\simeq{}$~few~keV. The following synopsis of this subject borrows heavily from these papers.

In our cases of interest, the nuclei are heavy compared with the electron mass or energy and compared with the emitted radiation. Therefore, we can think of the nuclei as static and we are essentially considering electrons scattering on the static Coulomb field of nuclei fixed in space. In this idealized situation, there are two sources of modification of the vacuum Coulomb field of a single nucleus. One is the screening provided by the electrons themselves which are highly degenerate and therefore ``difficult to polarize,'' whereas the other is the spatial correlation of the nuclei caused by their Coulomb repulsion. At low $T$, they actually arrange themselves in a body-centered cubic lattice and then are strongly correlated, not located independently at random relative positions. Strong correlations are more important in heavier WDs that have larger densities. The crystallization process was recently observed in the WD luminosity function of 0.9--$1.1\,M_\odot$ \cite{Tremblay:2019}, but plays no strong role for $0.6\,M_\odot$ WDs, let alone in RG cores. We will not have to worry about outright crystallization, yet we will have to worry about going to the intermediate correlation regime (``liquid phase'') beyond Debye screening.

In this Appendix we use the notation that the electron momenta in the scattering process are $\bk_1$ and $\bk_2$, whereas their momentum transfer is $\bk=\bk_1-\bk_2$. For very degenerate electrons, those able to scatter are at the Fermi surface with $|\bk_1|=|\bk_2|=k_{\rm F}$, the latter being the Fermi momentum. Therefore, as in the main text, $\bk^2=2k_{\rm F}^2(1-x_{12})$, where $x_{12}$ is the cosine of the angle between the in- and outgoing electron.

The deformation of a fully degenerate and homogeneous electron gas by an external test charge is governed by the Thomas-Fermi (TF) wave number that is
\begin{equation}
k_{\rm TF}=\left(\frac{4\alpha}{\pi}\,E_{\rm F}k_{\rm F}\right)^{1/2}
=\left(\frac{4\alpha}{\pi}\right)^{1/2}\left(m_e^2+k_{\rm F}^2\right)^{1/4} k_{\rm F}^{1/2}
=44.1\,{\rm keV}\,\left(1+0.641\,\rho_6^{2/3} \right)^{1/4}\,\rho_6^{1/6},
\end{equation}
where $\rho_6$ is $\rho$ in units of $10^6\,{\rm g}\,{\rm cm}^{-3}$ and we have assumed that $Y_e=Z/a=\frac{1}{2}$ electrons per baryon as will be the case in a medium consisting of $^4$He, $^{12}$C, or $^{16}$O. We also recall that \begin{equation}
    k_{\rm F}=\left(\frac{3\pi^2 Z\rho}{A m_u}\right)^{1/3}\bigg|_{A=2Z}
    =409\,{\rm keV}\,\rho_6^{1/3},
\end{equation}
The TF scale provides a Yukawa modification of the Coulomb potential or equivalently, in Fourier space, the squared Coulomb propagator gets modified as $|\bk|^{-4}\to (\bk^{2}+k_{\rm TF}^2)^{-2}$. Therefore, in a Coulomb integral, we should include
\begin{equation}
    \frac{1}{\bk^2}\to\frac{S_{\rm TF}(\bk)}{\bk^2}
\quad\hbox{where}\quad
S_{\rm TF}(\bk)=\left(\frac{\bk^2}{\bk^2+k_{\rm TF}^2}\right)^2,
\end{equation}
if degenerate electrons were the only source of screening.

The quasi-static nuclei (usually called ions in this context) require a different treatment. The electron scattering amplitudes from the ensemble of nuclei interfere coherently, where the interference term would average to zero if the targets were at random locations. Otherwise, the scattering process requires a ``static structure factor'' $S_{\rm i}(\bk)$ of the momentum transfer $\bk$, where the index i stands for ``ion.'' 
Therefore, in a Coulomb integral, we should include
\begin{equation}
    \frac{1}{\bk^2}\to  \frac{1}{\bk^2}\,S_{\rm i}(\bk).
\end{equation}
Averaging over directions of $\bk$, the structure factor is only a function of $|\bk|$. The limiting behavior of $S_{\rm i}(|\bk|)$ is 1 for large $|\bk|$ and it behaves as $|\bk|^2$ for small $|\bk|$. 

Notice that the squared matrix elements in Section~\ref{Sec:el_proton} involve a squared Coulomb propagator $1/|\bk|^4$. However, the phase-space integration over momentum transfers $\int d^3\bk=4\pi \int d|\bk|\,\bk^2$ introduces a factor $\bk^2$ in the numerator. So without screening, the rates would have a simple $1/\bk^2$ divergence that is logarithmic in the integrated rate, even though this may not be directly apparent from the expressions in Eqs.~\eqref{eq:kernel-p}--\eqref{eq:kernel-a}. Therefore, the $\bk^2$ scaling at low $|\bk|$ of the structure function is enough to moderate the divergence. As expected in a neutral medium, Coulomb scattering processes vanish in the forward direction (i.e.\ for vanishing momentum transfer). Unlike the TF prescription, that behaves as  $|\bk|^4$, one here does not modify the Coulomb field with a Yukawa factor and the resulting rates are not those that one would obtain from a screened Coulomb field, but we still refer to this modification as a screening effect. 

In a weakly correlated medium (sufficiently large $T$), the ion structure factor is given by the Debye formula
\begin{equation}
    S_{\rm i}(|\bk|)=\frac{\bk^2}{\bk^2+k_{\rm i}^2}
    \quad{\rm with}\quad
     k_{\rm i}^2=\frac{4\pi\alpha Z^2 n_{\rm i}}{T}.
\end{equation}
The mass density is $\rho=n_i A m_u$ with $A$ the atomic mass number (assuming only a single species) and $m_u=0.931\,{\rm GeV}$ the atomic mass unit. Therefore,
$Z^2 n_{\rm i}=Z (Z/A)\,\rho/m_u=(Z/2)\,\rho/m_u$ because in our media of interest, $Y_e=Z/A=1/2$. Therefore, numerically
\begin{equation}
     k_{\rm i}=\left(\frac{2\pi\alpha Z \rho}{m_uT}\right)^{1/2}=
     222\,{\rm keV}\,\left(\frac{Z_2\rho_6}{T_8}\right)^{1/2},
\end{equation}
where $T_8=T/10^8\,{\rm K}$ and $Z_2=Z/2$, corresponding to $^{4}$He, i.e., the reference conditions roughly correspond to a RG near helium ignition. (Recall that $10^8\,{\rm K}=8.6\,{\rm keV}$.) A dimensionless parameter that we often use in the main text is
\begin{equation}\label{eq:kappa2-def}
\kappa^2=\frac{k_{\rm i}^2}{2k_{\rm F}^2}
=\left(\frac{4\rho}{9\pi\,m_u}\right)^{1/3}\frac{Z\alpha}{T}
=0.147\,\frac{Z_2\rho_6^{1/3}}{T_8},
\end{equation}
again for $Z/A=1/2$ and the numerical factor corresponds roughly to a RG core.

As anticipated, the screening scale from degenerate electrons is much smaller compared to ion correlations. One way of including both effects is the prescription~\cite{Raffelt:1989zt}
\begin{equation}
    \frac{1}{|\bk|^4}\to \frac{S_{\rm i}(|\bk|)}{(|\bk|^2+k_{\rm TF}^2)^2}
 =\frac{1}{|\bk|^4}\,S_{\rm TF}(|\bk|)\,S_{\rm i}(|\bk|).   
\end{equation}
In Ref.~\cite{Nakagawa:1987pga} we can see for example in their Eqs.~(12) or (21) that they include electron screening with Jancovici's static dielectric function $\epsilon(k,0)$ that is given in their Eq.~(3). (Notice that they define the Thomas-Fermi scale with the non-relativistic formula where $E_{\rm F}=m_e$.) So they use $[1/\epsilon(k,0)]^2$ where we use $S_{\rm TF}(k)$. We have checked that the two expressions differ from each other only by a $k$-dependent factor of the order of $\alpha/\pi$ and therefore agree on the relevant level of perturbation theory.

\subsection{Strongly correlated plasma}\label{Sec:StrongCorrPlasma}

Beyond the Debye approximation, our picture is that of mobile ions immersed in an ``infinitely stiff'' electron background, i.e., a homogeneous neutralizing charge density. This is the traditional picture of a correlated one-component plasma. The usual ``plasma parameter'' to measure the strength of the correlations is the ratio of the ion-ion Coulomb interaction energy over their thermal kinetic energy, $\Gamma=Z^2\alpha/a_{\rm i}T$, where $a_{\rm i}$ is the ion-sphere radius given by $n_{\rm i}^{-1}=(4\pi/3)\,a_{\rm i}^3$, meaning that
\begin{equation}\label{eq:ion-sphere-radius}
    a_{\rm i}=\left(\frac{3 A m_u}{4\pi\,\rho}\right)^{1/3}\bigg|_{A=2Z}
    =\left(\frac{3 Z m_u}{2\pi\,\rho}\right)^{1/3}
    =\frac{1}{117~{\rm keV}}\,\frac{Z_6^{1/3}}{\rho_6^{1/3}},
\end{equation}
where $A=2Z$ and $Z_6=Z/6$ as for $^{12}$C. Numerically $\Gamma$ evaluates to 
\begin{equation}\label{eq:plasma-parameter}
\Gamma=\frac{Z^2\alpha}{a_{\rm i}T}
=\frac{Z^2\alpha}{T}\left(\frac{4\pi\rho}{3Am_u}\right)^{1/3}\bigg|_{A=2Z}
=\frac{Z^{5/3}\alpha}{T}\left(\frac{2 \,\pi \rho}{3m_u}\right)^{1/3}
=35.8\,\frac{Z_6^{5/3}\,\rho_6^{1/3}}{T_7}.
\end{equation}
It is also connected to our parameter $\kappa^2$ defined in Eq.~\eqref{eq:kappa2-def} through
\begin{equation}\label{eq:kappa-Gamma}
    \frac{\Gamma}{\kappa^2}=\left(\frac{3\pi^2}{2}\right)^{1/3}\!Z^{2/3}=8.11\,Z_6^{2/3}.
\end{equation}
Therefore, up to a factor, both quantities convey the same information. It is also worth noting that $k_{\rm i}^2 a_{\rm i}^2=3\Gamma$ and therefore the Debye structure factor can be expressed as
\begin{equation}\label{eq:Debye-Gamma}
    \frac{\bk^2}{\bk^2+k_{\rm i}^2}=\frac{|a_{\rm i}\bk|^2}{|a_{\rm i}\bk|^2+3\Gamma}.
\end{equation}
Strong correlations begin for $\Gamma\gtrsim 1$, corresponding to $\kappa^2\gtrsim0.12$ for $^{12}$C.

Overall we conclude that the Debye prescription is certainly good enough for RGs near helium ignition, whereas in WDs, especially toward the colder end of the luminosity function, the approximation is not necessarily sufficient. In the numerical simulations of axion emission in WDs \cite{MillerBertolami:2014rka}, the used emission rate actually included the screening prescription of Ref.~\cite{Nakagawa:1987pga} and therefore took account of strong correlations for $\Gamma>1$.

\begin{figure}[htp!]
    \centering
    \includegraphics[width=0.6\textwidth]{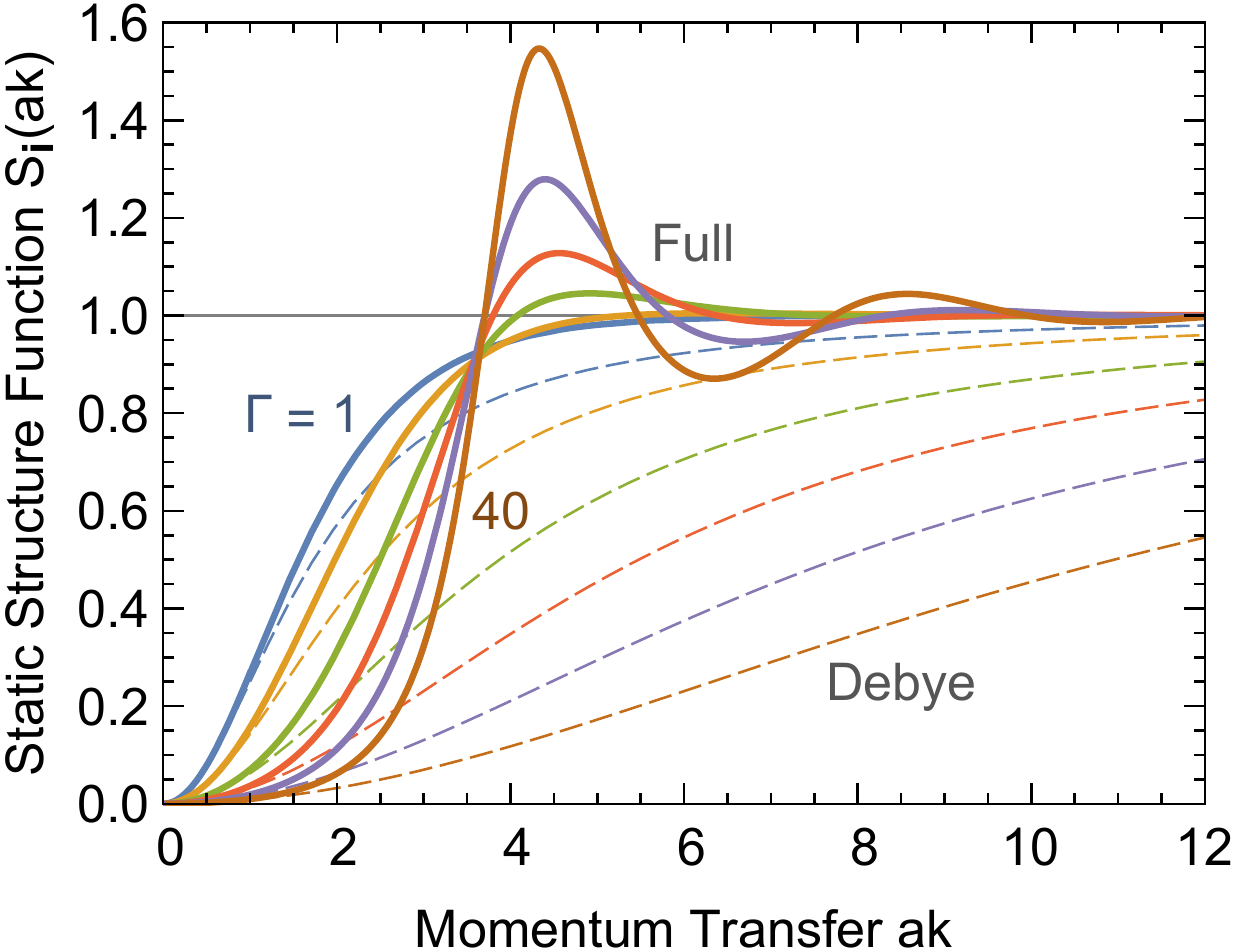}
    \caption{Ion-ion static structure factor for a one-component plasma (solid lines)  \cite{Itoh:1983,Ichimaru:1987PhR}. From upper left to lower right, following the colors from blue, orange, green, and so forth, these correspond to plasma parameters $\Gamma=1$, 2, 5, 10, 20, and 40. The dashed lines are the corresponding Debye structure functions of Eq.~\eqref{eq:Debye-Gamma} which asymptotically approach 1 for large $ak$, where screening is irrelevant, and agree with the full structure functions at small $ak$. The dimensionless momentum transfer is defined as $ak=|a_{\rm i}\bk|$ in terms of the ion-sphere radius $a_{\rm i}$ defined in Eq.~\eqref{eq:ion-sphere-radius}.}
    \label{fig:Structure-Function}
\end{figure}

In this regime one may use tabulated values for $S_{\rm i}(ak)$, where here $ak=|a_{\rm i}\bk|$ is a dimensionless momentum transfer in terms of the ion-sphere radius. Tabulations for a one-component plasma are found in Ref.~\cite{Itoh:1983} for $\Gamma=1$, 3, 6, 10, 20, 40, 80, 100, 125, and 160, whereas in Ref.~\cite{Ichimaru:1987PhR} for $\Gamma=2$, 5, 10, 20, 40, 80, 125, and 160, both going back to Ichimaru and collaborators. In the regions of overlap, both sets are practically identical. In Fig.~\ref{fig:Structure-Function} we show the numerical results (solid lines) from these data, i.e., $S_{\rm i}(ak)$ and compare them with the Debye approximation (dashed lines). We show the results for $\Gamma=1$, 2, 5, 10, 20 and 40, from upper left to lower right, with colors blue, orange, green, and so forth.

To estimate quantitatively the impact of ion-ion correlations, we may imagine that the rates are expanded in powers of $\beta_{\rm F}$, where actually only even powers of $\beta_{\rm F}$ appear. In the absence of screening effects, we need integrals of the form of 
Eq.~\eqref{eq:Fi-def}, i.e., we need
\begin{equation}
    F_0=\int_{-1}^{+1}dx_{12}\,(1-x_{12})^n\,\frac{1}{(1-x_{12})}.
\end{equation}
Here, $n=0$ is the simple Coulomb integral that we need for the $\beta_{\rm F}=0$ limit, wheres we need also $n=1$ if we go to the next order $\beta_{\rm F}^2$.

To include the static structure function, we observe that the momentum transfer $|\bk|$ ranges from 0 to $2k_{\rm F}$ and that $S_{\rm i}(ak)$ is provided in terms of the dimensionless momentum transfer $ak=|a_{\rm i}\bk|$. Therefore, under the Coulomb integrals we must include the factor
\begin{equation}
S_{\rm i}\Bigl[a_{\rm i}k_{\rm F}\sqrt{2(1-x_{12})}\Bigr],  
\end{equation}
where
\begin{equation}
    a_{\rm i}k_{\rm F}=\left(\frac{9\pi Z}{4}\right)^{1/3}=\left(\frac{63\pi}{4}\right)^{1/3}Z_6^{1/3}
    = 3.49\,Z_6^{1/3},
\end{equation}
where we have chosen the ion charge 6 as a reference value for $^{12}$C in a WD. To include Thomas-Fermi screening, we need the further factor
\begin{equation}
\left(\frac{1-x_{12}}{1-x_{12}+\kappa^2_{\rm TF}}\right)^2,  
\end{equation}
where
\begin{equation}
    \kappa^2_{\rm TF}=\frac{k_{\rm TF}^2}{2k_{\rm F}^2}=\frac{2\alpha}{\pi}\,\frac{m_e}{k_{\rm F}}\left(1+\frac{k_{\rm F}^2}{m_e^2}\right)^{1/2}=5.80\times10^{-3}\,\frac{\sqrt{1+0.641\,\rho_6^{2/3}}}{\rho_6^{1/3}}.
\end{equation}
Therefore, overall the Coulomb integrals are
\begin{equation}
    F_n=\int_{-1}^{+1}dx_{12}\,(1-x_{12})^n\,\left(\frac{1-x_{12}}{1-x_{12}+\kappa^2_{\rm TF}}\right)^2\, 
    \frac{S_{\rm i}\Bigl[a_{\rm i}k_{\rm F}\sqrt{2(1-x_{12})}\Bigr]}{(1-x_{12})}.
\end{equation}
In the Debye limit ($\Gamma\ll1$), and ignoring the TF term ($\kappa_{\rm TF}^2=0$), the first two integrals are
\begin{eqnarray}
  F_0(\Gamma)\big|_{\rm Debye}&=&\log\left[1+\frac{4 (a_{\rm i}k_{\rm F})^2}{3\Gamma}\right],
  \label{eq:G0}
  \\[1.5ex]
  F_1(\Gamma)\big|_{\rm Debye}
    &=&2\left[1-\frac{3\Gamma}{4 a_{\rm i}k_{\rm F}}\log\left(1+\frac{4 a_{\rm i}k_{\rm F}}{3\Gamma}\right)\right].
\end{eqnarray}
The integrals can also be done including the TF term, but the expressions are too complicated to be illuminating.

\begin{figure}[htp!]
    \centering
    \hbox to\textwidth{\hfil \includegraphics[width=0.45\textwidth]{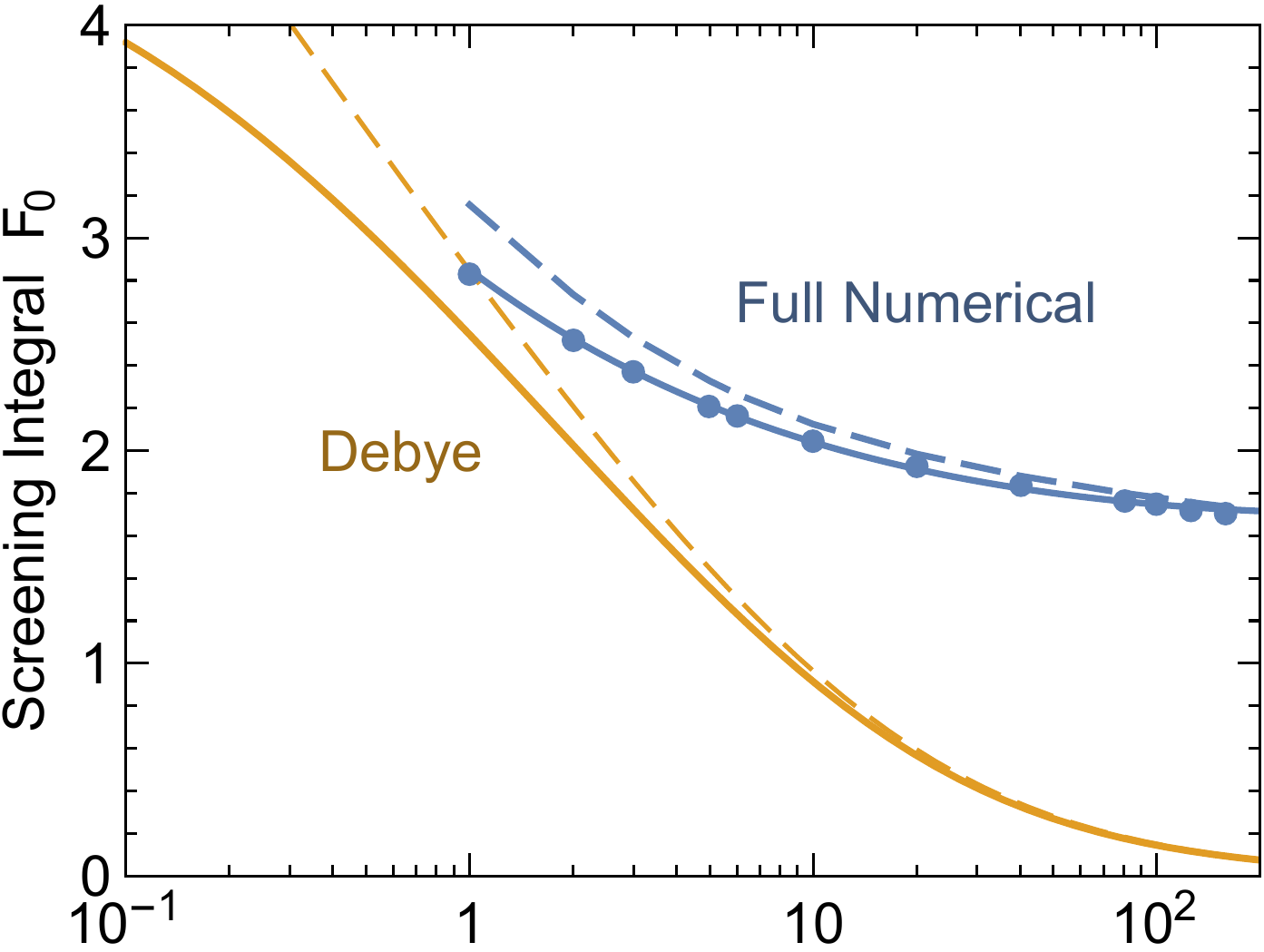}\hfil}
    \vskip2pt
   \hbox to\textwidth{\hfil \includegraphics[width=0.45\textwidth]{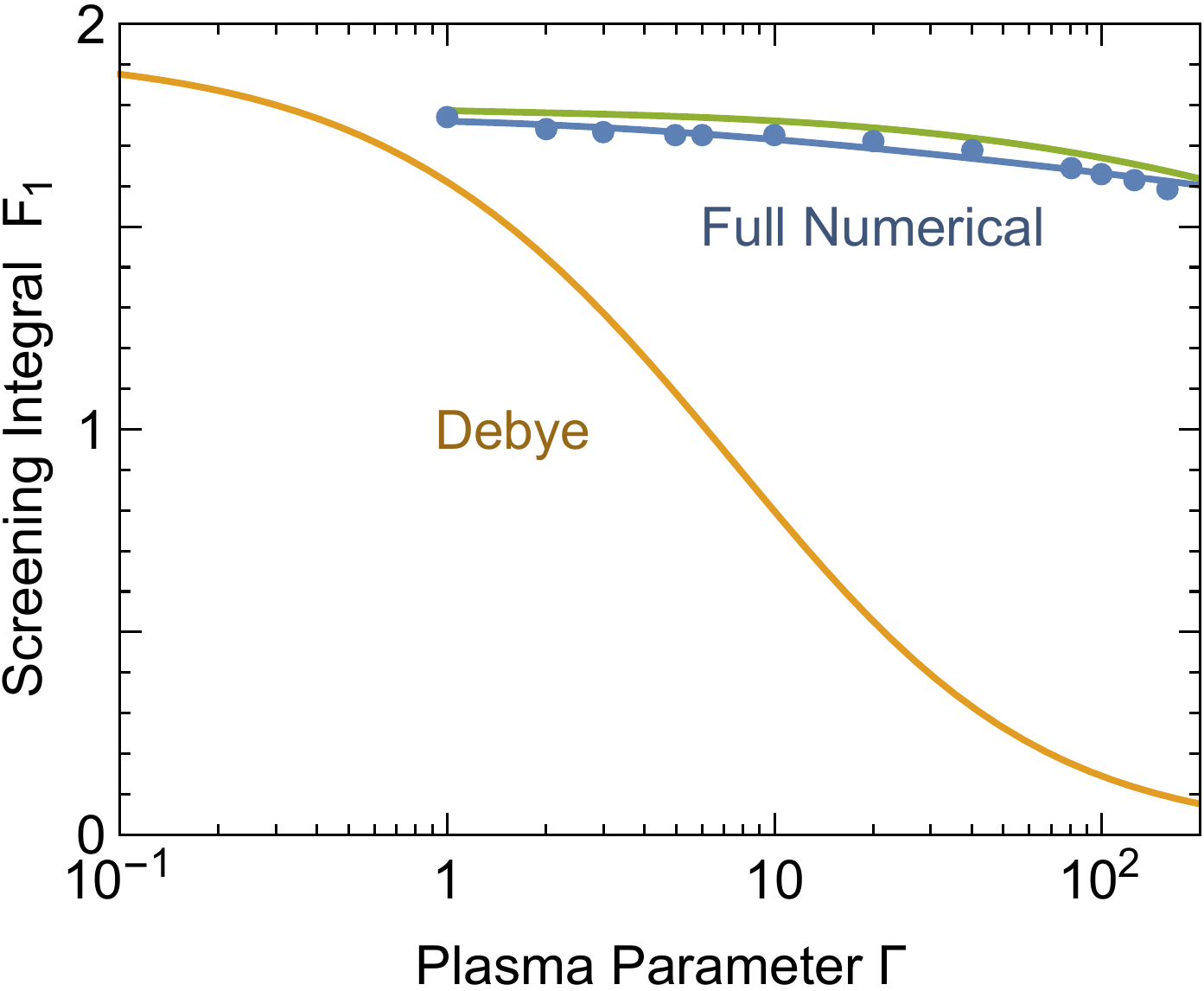}\hfil}
    \caption{Numerical screening integrals $F_0(\Gamma)$ and $F_1(\Gamma)$ for a one-component plasma with $Z=6$ (carbon) as blue dots, including both static ion-ion correlations and Thomas-Fermi screening by degenerate electrons. The solid blue lines are our fit functions Eqs.~\eqref{eq:FitScreening0} and~\eqref{eq:FitScreening1}. In the upper panel, the dashed blue line is the same without TF. In the lower panel, the green line is the fit function of Itoh et al. \cite{Itoh:1983}. The solid orange line is the Debye result, dashed without TF.}
    \label{fig:FullScreening}
\end{figure}
 
In Fig.~\ref{fig:FullScreening} we show the Coulomb integrals $F_0(\Gamma)$ and $F_1(\Gamma)$, assuming $Z=6$ (carbon) and a density of $\varrho=10^6~{\rm g}~{\rm cm}^{-3}$. The blue dots are the numerical integrals for the tabulated static structure functions and also include the TF screening by degenerate electrons. We also show the Debye result, and in the upper panel also the result in the absence of TF screening (dashed lines). We also show the fit functions (blue lines)
\begin{subequations}
\begin{eqnarray}\label{eq:FitScreening0}
F_0(\Gamma)\big|_{\rm fit}&=&+1.77\,\Gamma^{-0.33} + 1.08\,\Gamma^{+0.05},
\\[1ex]
\label{eq:FitScreening1}
F_1(\Gamma)\big|_{\rm fit}&=&-3.82\,\Gamma^{-0.1~}+ 5.58\,\Gamma^{-0.07}.
\end{eqnarray}
\end{subequations}
Itoh et al.~\cite{Itoh:1983} have provided analytic fit functions, where our $F_0$ is what they call $2\langle S_{-1}\rangle$ and $F_1$ is what they call $4\langle S_{+1}\rangle$. For $F_0$, their fit function virtually overlays with ours and the agreement is very good. For $F_1$ we show their fit function as a green line. The agreement is slightly worse, but still, as $F_1$ would appear together with a factor $\beta_{\rm F}^2$, the overall error would be small.

\subsection{Full numerical integration for carbon and oxygen}\label{Sec:FullNumeric}

Clearly the emission rates will be significantly larger in a WD than predicted by a naive application of Debye screening. Moreover, the non-relativistic expansion using only the $\beta_{\rm F}=0$ limit is somewhat rough for WD conditions. Therefore, we consider the full expressions for baryophilic scalar, leptophilic scalar, and axion emission through their electron coupling. Our data can be represented by a fit function of the form
\begin{equation}\label{eq:FitFunction-App}
F_{\rm fit}(\rho,\Gamma)=A(\rho)\,\Gamma^{-0.37} + B(\rho)\,\Gamma^{+0.03}.
\end{equation}
For all cases, the coefficient functions are found to be well fitted by the functional form
\begin{subequations}\label{Eq:FunctionalFormNew}
\begin{eqnarray}
A(\rho)&=& a_{0} + a_{1} x + \frac{a_{2}}{(8-x)} + \frac{a_{3}}{(8-x)^2},
\\[1ex]
B(\rho)&=& b_{0} + b_{1} x + \frac{b_{2}}{(8-x)} + \frac{b_{3}}{(8-x)^2},
\end{eqnarray}    
\end{subequations}
where $x=\log_{10}(\varrho)$ with $\rho$ in units of ${\rm g}/{\rm cm}^3$. The numerical coefficients are different for different atomic charge $Z$ and different bosons
(Table~\ref{tab:FitParametersNew}). The fit applies to the range $4\leq x\leq 7$ and $1\leq\Gamma\leq 160$, where it is typically good at the few~\% level. For $x \gtrsim 5$, the fit works better than 1\%.
\begin{table}[H]
\caption{Coefficients for the fit functions of Eqs.~\eqref{Eq:FunctionalFormNew}.}
    \label{tab:FitParametersNew}
    \vskip6pt
    \centering
    \begin{tabular}{lllll}
    \hline
    \hline
        & $a_0$ & $a_1$ & $a_2$ & $a_3$ \\
        & $b_0$ & $b_1$ & $b_2$ & $b_3$ \\    
         \hline
         \multicolumn{5}{l}{Carbon\vphantom{$\Big[$} ($Z=6$)}\\[1ex] 
        \quad Baryon & +0.665 & $-$0.244 & +6.173 & +0.713  \\
                  & +1.345  & $-$0.201 & +3.00 & +0.057 \\
        \quad Lepton & +0.567 & $-$0.217 & +6.413 & $-$0.543    \\ & +1.214  & $-$0.004 & +0.327 & $-$0.231 \\
        \quad Axion & +0.248 & +0.306 & $-$1.145 & +0.393   \\ & +1.293  & +0.152 & $-$2.918 & +1.200 \\[0.1ex] 
        \hline
        \multicolumn{5}{l}{Oxygen\vphantom{$\Big[$} ($Z=8$)}\\[1ex] 
        \quad Baryon & +0.560 & $-$0.229 & +6.161 & +0.749   \\ & +1.492  & $-$0.240 & +3.660 & +0.096 \\
        \quad Lepton & +0.468 & $-$0.199 & +6.305 & $-$0.368    \\ & +1.338  & $-$0.040 & +1.093 & $-$0.452 \\
        \quad Axion & +0.133 & +0.325 & $-$1.155 & +0.410   \\ & +1.401  & +0.169 & $-$3.006 & +1.213 \\
        \hline
    \end{tabular}
\end{table}

\begin{figure}[H]
    \hbox to\textwidth{\hfil \includegraphics[width=0.48\textwidth]{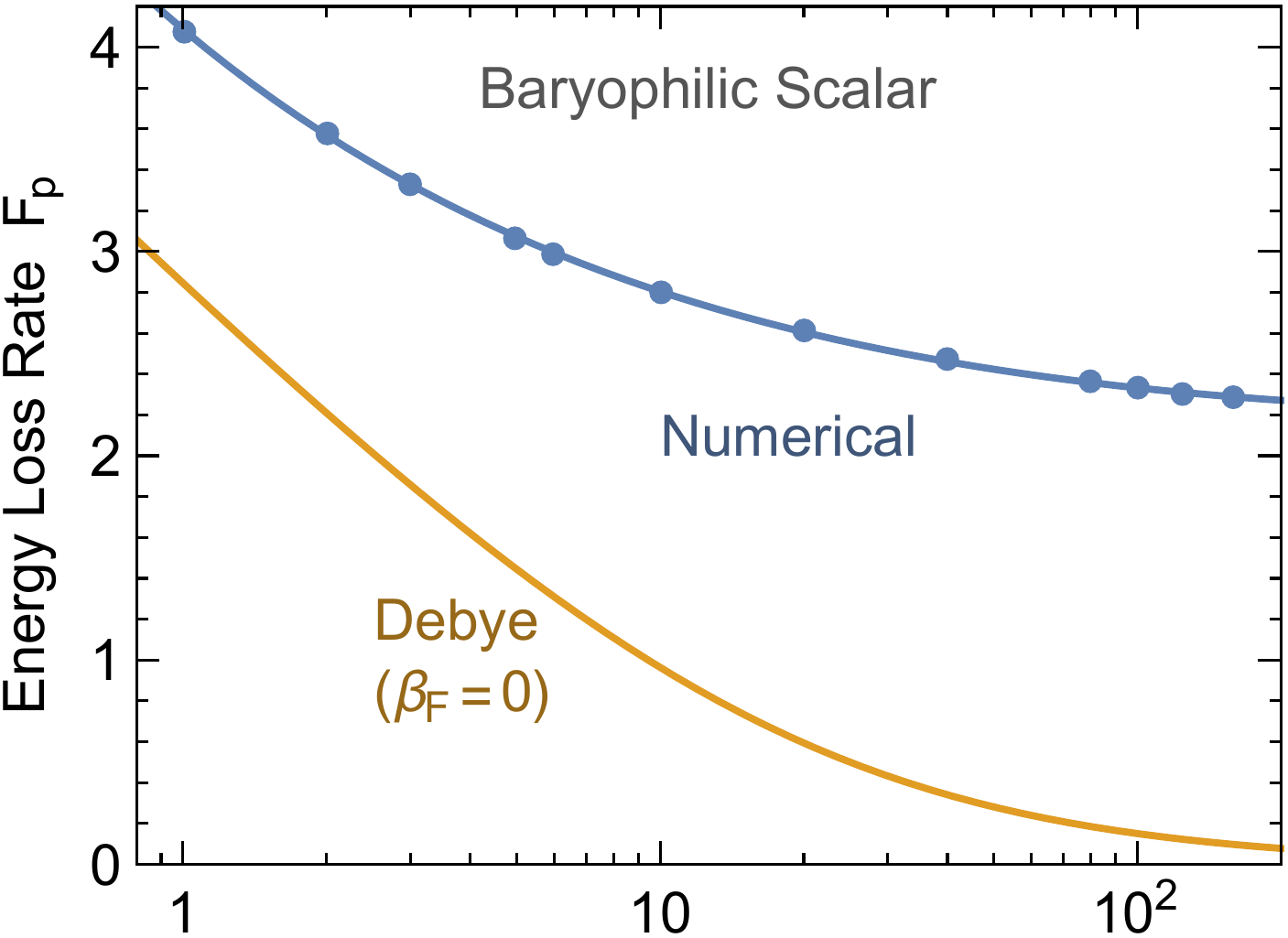}\hfil}
    \vskip5pt
   \hbox to\textwidth{\hfil \includegraphics[width=0.48\textwidth]{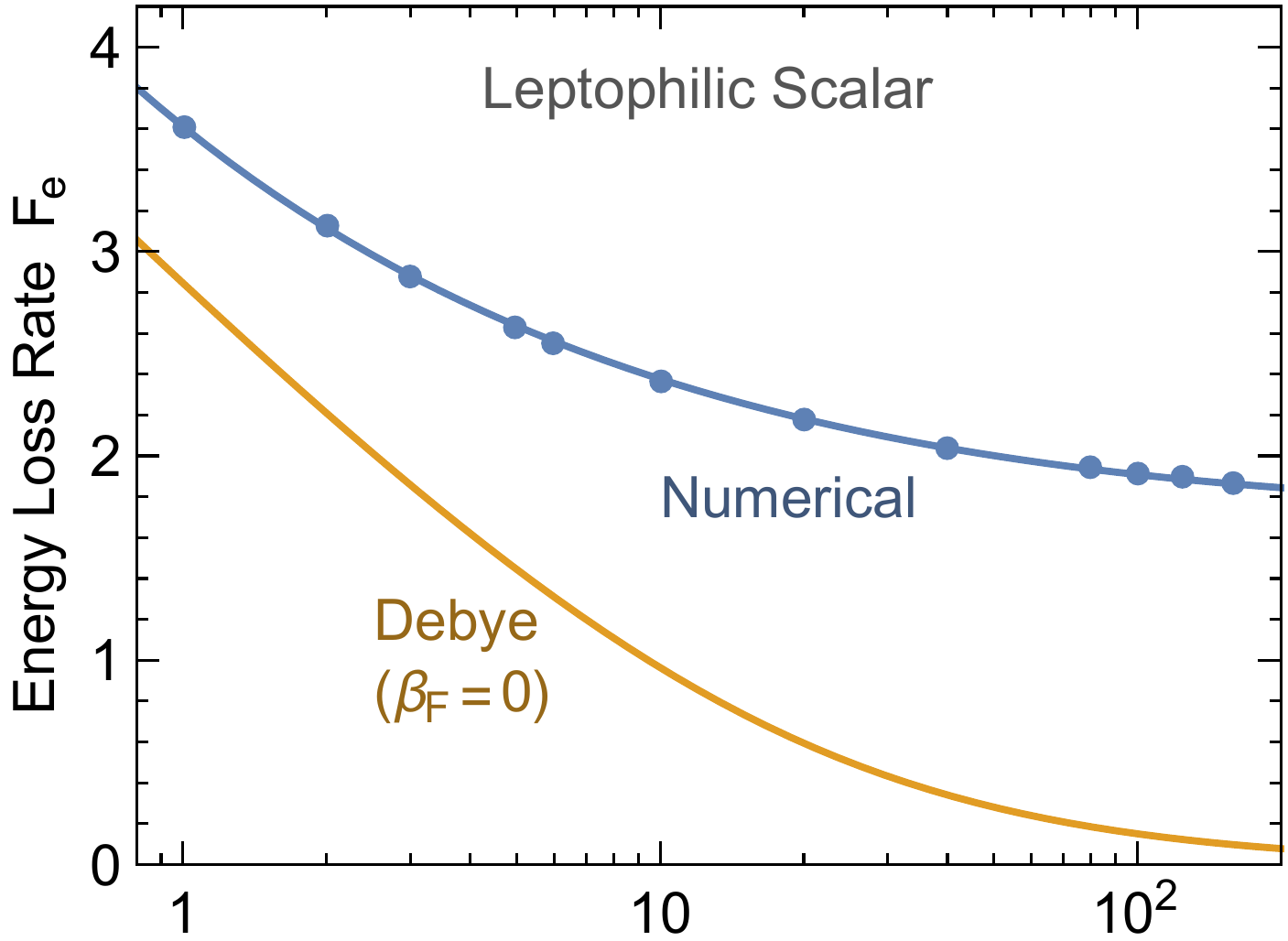}\hfil}
   \vskip5pt
   \hbox to\textwidth{\hfil \includegraphics[width=0.48\textwidth]{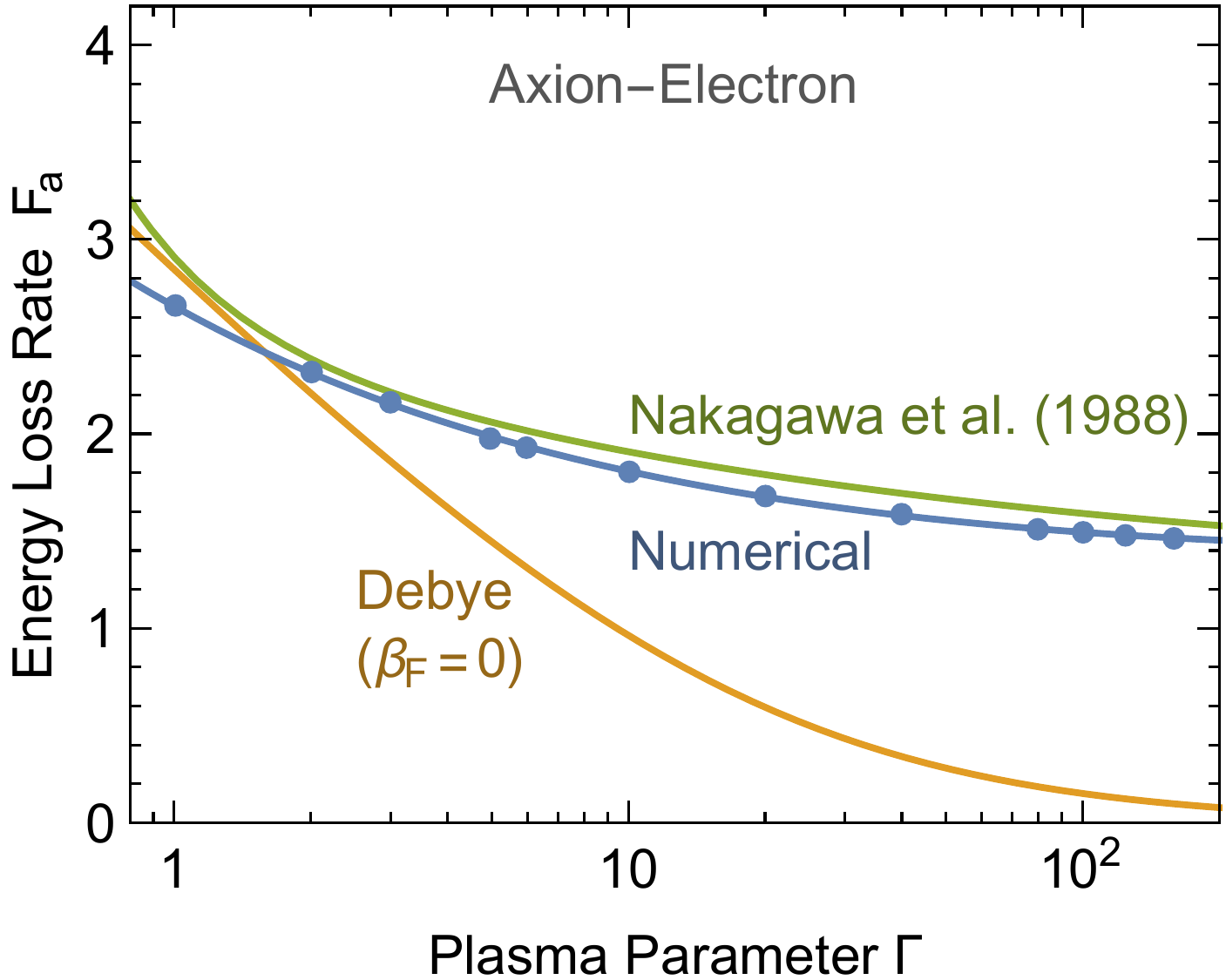}\hfil}
    \caption{Energy-loss rate for $\rho=10^6~{\rm g}~{\rm cm}^{-3}$ and $Z=6$ (carbon) for different bosons. The curve ``Debye'' given by Eq.~\eqref{eq:G0} is the same in all panels and, apart from global factors, is the emission rate for $\beta_{\rm F}=0$ and includes only ion-ion correlations in the Debye limit. The data points come from numerical integrations with the full ion-ion correlation function and include Thomas-Fermi screening by the electrons. The blue lines are the fit function of Eq.~\eqref{Eq:FunctionalFormNew}. For axions, the green curve is the fit function of Nakagawa et al.~\cite{Nakagawa:1988rhp} that is claimed to be accurate to better than 20\%, but much better for this example.}
    \label{fig:EmissionRates}
\end{figure}

In order to be concrete and to illustrate the quality of our fitting formulas, we consider once more WD conditions with the density $\rho=10^6~{\rm g}~{\rm cm}^{-3}$ and $Z=6$ (carbon). In this case, the various physical parameters are: Fermi momentum $k_{\rm F}=409$~keV. Thomas-Fermi wave number: $k_{\rm TF}=51.0$~keV. Velocity at Fermi surface: $\beta_{\rm F}=0.625$. Screening scale from ions: $k_{\rm i}=1216\,{\rm keV}/T_7$, where $T_7=T/10^7$~K. Ion-sphere radius: $a_{\rm i}^{-1}=117$~keV. Plasma parameter: $\Gamma=35.8/T_7$. In Fig.~\ref{fig:EmissionRates} we show the emission rates for our three generic scalar boson models.

  In every panel, we show the ``naive Debye'' rate as an orange line. This is simply the Coulomb integral $F_0$ with the inclusion of only the static ion-ion correlation. The full numerical integration, including the Thomas-Fermi screening, for the available tabulations of the ion-ion correlation are shown as blue dots. The results of the fitting formulas Eq.~\eqref{eq:FitFunction-App} are shown as blue lines. For axions, analytic fitting formulas were already provided by Nakagawa et al.~\cite{Nakagawa:1988rhp} (green line), which agree with our results at their claimed level of accuracy of better than~20\%.

\bibliographystyle{bibi}
\bibliography{Biblio}

\end{document}